\documentclass{article}

\usepackage{PRIMEarxiv}

\usepackage{natbib}
\usepackage[utf8]{inputenc} 
\usepackage[T1]{fontenc}    
\usepackage{hyperref}       
\usepackage{url}            
\usepackage{enumitem}
\usepackage{booktabs}       
\usepackage{amsfonts}       
\usepackage{amsthm}
\usepackage{amsmath}
\usepackage{nicefrac}       
\usepackage{microtype}      
\usepackage{lipsum}
\usepackage{fancyhdr}       
\usepackage{graphicx}       
\usepackage{subcaption} 
\usepackage{caption}    
\usepackage{algorithm}
\usepackage{algorithmic}

\graphicspath{{media/}}     
\makeatletter

\title{Heterogeneity-Aware Federated Causal Inference Leveraging Effect-Measure Transportability
}

\author{
  Siqi Cao \\
  Department of Statistics\\
  North Carolina State University \\
  Raleigh, North Carolina 27695, U.S.A.\\
  \texttt{scao7@ncsu.edu} \\
   \And
  Shu Yang \\
  Department of Statistics \\
  North Carolina State University \\
  Raleigh, North Carolina 27695, U.S.A.\\
  \texttt{syang24@ncsu.edu} \\
}

\AtBeginDocument{%
  \makeatletter
  \newcommand{\assumpsup}[1]{\gdef\theassumptionsup{\textsuperscript{(#1)}}}
  \newcommand{\clearassumpsup}{\gdef\theassumptionsup{}}
  \clearassumpsup

  \newcommand{\condsup}[1]{\gdef\theconditionsup{\textsuperscript{(#1)}}}
  \newcommand{\clearcondsup}{\gdef\theconditionsup{}}
  \clearcondsup

  \makeatother
}

\newtheorem{theorem}{Theorem}

\newtheorem{lemma}{Lemma}
\newtheorem{corollary}{Corollary}
\newtheorem{assumption}{Assumption}
\newtheorem{condition}{Condition}

\theoremstyle{definition}

\theoremstyle{remark}
\newtheorem{remark}{Remark}

\begin{document}
\maketitle

\begin{abstract}
Federated learning of causal estimands offers a powerful strategy to improve estimation efficiency by leveraging data from multiple study sites while preserving privacy. Existing literature has primarily focused on the average treatment effect using single data source, whereas our work addresses a broader class of causal measures across multiple sources. We derive and compare semiparametrically efficient estimators under two transportability assumptions, which impose different restrictions on the data likelihood and illustrate the efficiency-robustness tradeoff. This estimator also permits the incorporation of flexible machine learning algorithms for nuisance functions while maintaining parametric convergence rates and nominal coverage. To further handle scenarios where some source sites violate transportability, we propose a Post-Federated Weighting Selection (PFWS) framework, which is a two-step procedure that adaptively identifies compatible sites and achieves the semiparametric efficiency bound asymptotically. This framework mitigates the efficiency loss of weighting methods and the instability and computational burden of direct site selection in finite samples. Through extensive simulations and real-data analysis, we demonstrate that our PFWS framework achieves superior variance efficiency compared with the target-only analyses across diverse transportability scenarios.
\end{abstract}


\section{Introduction}
In many applications, the same treatment is administered to populations observed in distinct environments, yet the data remain stored separately. When one study is too small to yield precise estimates, it is natural to seek efficiency gains by integrating information across sources. Direct pooling of individual-level data, however, is often precluded by legal constraints, privacy concerns, or proprietary barriers, thereby motivating federated approaches that rely only on site-level aggregates. Such integration is further complicated by differences in the causal measures of interest, heterogeneity in covariate and outcome distributions, and the presence of incompatible sources that, if indiscriminately combined, may invalidate inference.

Traditional approaches, such as meta-analysis, aggregate evidence from independent studies while accounting for study-level heterogeneity \citep{hartung2011statistical}. Extending beyond this classical framework, recent work in causal inference has sought to integrate data across heterogeneous sites while addressing violations of transportability, a notion often referred to under different but essentially equivalent terminologies, such as exchangeability, invariance and homogeneity. Examples include selective borrowing framework based on individual-level data \citep{gao2025improving}, test-then-pool procedures \citep{yang2023elastic, gao2023pretest}, and bias function modeling \citep{yang2025data, wu2022integrative, cheng2024inference}. Some recent work has shifted toward federated settings, motivated by privacy constraints that prevent the pooling of individual-level data. Within this paradigm, robust estimators \citep{han2023multiply,xiong2023federated} and site-adaptive weighting schemes \citep{han2025federated,liu2024multi} have been proposed to achieve efficient estimation by sharing only site-level aggregates.

Recent studies have increasingly emphasized leveraging data heterogeneity across multiple sites to improve estimation of Average Treatment Effect (ATE) estimation, typically represented by the difference in means \citep{han2023multiply, xiong2023federated, han2025federated}. However, established clinical reporting standards, notably the CONSORT guidelines \citep{moher2010consort}, emphasize the importance of presenting treatment effects through both absolute (e.g., risk difference) and relative (e.g., risk ratio)
metrics to provide a more comprehensive assessment of treatment effects. The choice of scale has substantive implications: empirical evidence shows that effect framing significantly influences clinical decision-making \citep{naylor1992measured, forrow1992absolutely}, and treatment effect heterogeneity itself may be scale-dependent \citep{rothman2024epidemiology}. Motivated by these considerations, we study a broad class of causal measures, with a primary focus on the Risk Ratio (RR), to illustrate the our methodology and its properties.

Some recent work addresses the integration of multiple data sources but often under strong transportability assumptions. \citet{colnet2023risk} establishes identifiability of the causal estimand under a weaker condition by assuming the transportability of the treatment effect. In contrast to most approaches that assume transportability of the conditional outcome, we show that estimators based on this weaker assumption yield larger semiparametric bound but offer greater robustness due to fewer restrictions on the data likelihood function. Only a few studies have considered deriving efficient influence function (EIF) under this weaker assumption: \citet{su2024efficient} focuses solely on the mean difference, and \citet{li2024efficient} addresses the case with a single external data set. Our work fills these gaps by characterizing the EIF for a broader class of causal measures across multiple sites under this weaker assumption, an open problem explicitly noted in \citet{su2024efficient}. Using the RR as an illustrative example, we derive its explicit EIF and highlight its desirable properties, including multiple robustness, rate robustness, and local efficiency. We also compare this EIF with the EIF derived under the stricter transportability assumption commonly used in the literature.

As the complexity of data integration across multiple sites increases, addressing violations of key assumptions, such as transportability, has become crucial. Some recent work, for example \citet{gao2025improving}, considers individual selection method, but it is not developed in a federated setting. Most existing federated estimators employ a weighted framework to mitigate such violations \citep{liu2024multi, han2025federated}. However, weighting approaches often sacrifice statistical efficiency in finite samples, as the weights for incompatible sites cannot always be shrunk to zero. In contrast, directly applying the EIF-based estimator to the subset of sites that satisfy the necessary identification assumptions and conditions yields the smallest achievable semiparametric variance bound under this case. However, selecting the appropriate subset of sites becomes increasingly challenging as the number of sites grows, due to the exponential increase in possible combinations. To address this, we propose the Post-Federated Weighting Selection (PFWS) framework, which integrates federated weighting with a data-adaptive selection procedure to identify compatible sites. PFWS effectively alleviates the efficiency loss of weighted method under finite samples and attains the semiparaemtric efficiency bound asymptotically.

\section{Notations, assumptions and preliminaries}
\label{notation}
Suppose we have $K+1$ sites, where $K$ is finite. Define the site index set as $[[K]]=\{0,1,2,\cdots,K\}$. Each site $k\in[[K]]$ contains $n_k$ observations. The total sample size is $n=\sum_{k=0}^Kn_k$. Each individual is characterized by the random tuple $V=\{Y(1),Y(0),X,A,S\}\in\mathbb R\times\mathbb R\times\mathcal{X}\times\{0,1\}\times\{0,\cdots,K\}$. Here, $Y(a)$ for $a\in\{0,1\}$ denotes the potential outcome under treatment $a$, $X$ is a $p$-dimensional covariate vector with $\mathcal{X}\subset{\mathbb R}^p$, $A$ is the binary treatment assignment, and $S$ is the site indicator, where $S=k$ denotes subjects from site $k$. 

The causal measure of interest is defined as a function $m:D_m\rightarrow\mathbb R$ for some domain $D_m\subset\mathbb R^2$:
\begin{equation}
\label{cm}
    \psi^{(k)}=m\left(\mathbb E\{Y(0)|S=k\},\mathbb E\{Y(1)|S=k\}\right),\quad k\in[[K]].
\end{equation}
Without loss of generality, let site $0$ be our target site. The causal estimand of interest is then the site-specific effect $\psi^{(0)}$ under the chosen causal measure.

For each site $k\in[[K]]$, let the conditional causal measure be
$\tau^{(k)}(x)=m(\mathbb E\{Y(0)|X=x,S=k\},\mathbb E\{Y(1)|X=x,S=k\})$, the site-specific propensity score be
$\pi^{(k)}(x)=\operatorname{pr}(A=1|X=x,S=k)$, and the conditional mean outcome under treatment $a\in\{0,1\}$ be
$\mu_a^{(k)}(x)=\mathbb E\{Y(a)|X=x,S=k\}$. To quantify the differences in covariate distributions between the target site $0$ and a site $k$, we define the density ratio as $q^{(k)}(x)=\operatorname{pr}(S=0|X=x)/\operatorname{pr}(S=k|X=x)$.

To proceed, we begin by making the following standard assumptions. For each site, we assume unconfoundedness and positivity, which are commonly imposed in the causal inference literature to ensure the identifiability of causal measures. While these are idealized, we later consider scenarios in which some of them can be violated. 

\assumpsup{k}
\begin{assumption}[Unconfoundedness for site $k$]
\label{unconf}
$\{Y(0),Y(1)\}\perp A|X,S=k$.
\end{assumption}
\clearassumpsup

\assumpsup{k}
\begin{assumption}[Positivity for site $k$]
\label{posi}
For $\eta>0$, the propensity score function holds $\eta<\pi^{(k)}(x)<1-\eta$ for all $x$ such that $\operatorname{pr}(X=x,S=k)>0$.
\end{assumption}
\clearassumpsup


To establish rigorous identification of the causal effect measure $\psi$ (defined in Equation~\ref{cm}), we impose a structural constraint ensuring that the mapping preserves distinct causal information.

\clearassumpsup
\begin{assumption}[Injectivity]
\label{inje}
Let $\psi$ be the causal measure defined in Equation~\ref{cm}. $D_m\subset\mathbb R^2$ is the domain of function $m(\cdot,\cdot)$. For each $z$ that appears as the first coordinate in some pair of $D_m$, define $D_m^{(2)}(z):=\{z'\in\mathbb R:(z,z')\in D_m\}$ and $g_z:D_m^{(2)}(z) \rightarrow \mathbb{R}, z' \mapsto m(z, z')$. We assume that for every such $z$, the function \( g_z \) is injective, that is, for $z_1'\neq z_2'$, we have $g_z(z_1')\neq g_z(z_2')$.
\end{assumption}

\begin{remark}
    Assumption \ref{inje}, which states that $g_z$ is injective, is mild. If this is violated, two different values of $\mathbb E\{Y(1)|X\}$ could correspond to the same conditional causal measure for a given baseline $\mathbb E\{Y(0)|X\}$.
\end{remark}
 
Under Assumption \ref{inje}, the binary nature of $A$ allows the response $Y$ to be decomposed into two components: the baseline level and the treatment-induced modification. This decomposition is general and does not reply on any parametric assumptions. The following lemma formalizes this result:
\begin{lemma}
\label{lemma}
Let $\psi$ be a causal measure defined in Equation~\ref{cm} that satisfies Assumption \ref{inje}. Then, for every distribution of \( (Y(0), Y(1)) \mid X \), there exist two unique functions \( \mu_0, \tau : \mathcal{X} \to \mathbb{R} \) such that, for all \( x \in \mathcal{X} \) where $\left(\mathbb{E}\{Y(0) \mid X = x\}, \mathbb{E}\{Y(1) \mid X = x\}\right) \in D_m$, we have $
\mathbb{E}\{Y(0) \mid X = x\} = \mu_0(x) \quad $ and $\quad \mathbb{E}\{Y(1) \mid X = x\} = g_{\mu_0(x)}^{-1}(\tau(x)).
$
\end{lemma}

\begin{remark}
    For each covariate value $x$, the baseline level is captured by $\mu_0(x)=\mathbb E\{Y(0)|X=x\}$. The treatment-induced modification is captured by $\tau(x)$, which, through the injective mapping $g_{\mu_0(x)}^{-1}$, uniquely determines the expected outcome under treatment, $\mathbb E\{Y(1)|X=x\}$.
\end{remark}
Applying Lemma \ref{lemma} to the Rish Difference (RD) and RR leads to the following Corollary.

\begin{corollary}
    In the framework of Lemma \ref{lemma}, we have $\mu_0(X)=\mathbb E\{Y(0)|X\}$.
    \begin{enumerate}[label=(\roman*)]
        \item Consider RD. Let $g_z(z')=z'-z$ and $g_z^{-1}(z')=z'+z$, then
        $\mathbb E\{Y(1)|X\}=\tau(X)+\mu_0(X).$

        \item Consider RR. Let $g_z(z')=z'/z$ and $g_z^{-1}(z')=z'z$, then
        $\mathbb E\{Y(1)|X\}=\tau(X)\mu_0(X).$
    \end{enumerate}
\end{corollary}

    The definition of the function $g_z(z')$ is essential for deriving a general procedure to calculate the EIF for any causal measure that satisfies Assumption \ref{inje}. We demonstrate this in the Supplementary Material.

To borrow the information from the source sites, additional assumptions are required. For causal measures satisfying Assumption \ref{inje}, two identification strategies exist according to \citet{colnet2023risk}: (\romannumeral1) assuming transportability of the local effect measure, or (\romannumeral2) assuming transportability of the conditional outcomes. These two strategies correspond to the following conditions:

\condsup{k}
\begin{condition}[Transportability of treatment effect for source site $k$]
\label{con1}
For source site $k\in[[K]]\backslash\{0\}$, $\tau^{(0)}(X)=\tau^{(k)}(X)$.
\end{condition}
\clearassumpsup

\condsup{k}
\begin{condition}[Transportability of conditional outcome for source site $k$]
\label{con2}
For source site $k\in[[K]]\backslash\{0\}$, $\mu_a^{(0)}(X)=\mu_a^{(k)}(X)\text{ for }a\in\{0,1\}$.
\end{condition}
\clearassumpsup

Both Condition \ref{con1} and Condition \ref{con2} enable the identification of $\psi^{(0)}$ using data from site $k$, regardless of the chosen causal measure. Based on each condition, we can construct semiparametrically efficient estimators separately. Notably, Condition \ref{con2} is stricter than Condition \ref{con1}, implying that any site $k$ satisfying Condition \ref{con2} also satisfies Condition \ref{con1}. For notational convenience, we write Condition $1^{(a:b)}$ to indicate that Condition $1^{(k)}$ holds for all $k=a,\dots,b$. The same superscript notation applies to other conditions and assumptions.

However, in practice, the validity of these conditions across sites is unknown. This paper therefore has two main objectives: (\romannumeral1) assuming either Condition $1^{(1:K)}$ or Condition $2^{(1:K)}$, we develop a semiparametrically efficient and multiply robust estimator that leverages the data from source sites to improve estimation efficiency ($\S$\ref{MR}); (\romannumeral2) Acknowledging the potential violation of Assumption \ref{unconf}, \ref{posi}, or Condition \ref{con1} for some sites $k\in[[K]]\backslash\{0\}$, in $\S$\ref{fed}, we propose a federated selection procedure to identify the sites satisfying these assumptions and condition, thereby ensuring valid application of the estimator developed in $\S$\ref{MR}.

\section{Semiparametric efficient estimators}
\label{MR}
\subsection{Semiparametric efficient point estimators under ideal situation}
\label{est}
In $\S$\ref{MR}, we assume all source sites are transportable, that is, Assumptions $1^{(0:K)}$ and $2^{(0:K)}$ hold, and either Condition $1^{(1:K)}$ or Condition $2^{(1:K)}$ is satisfied. We then focus on the RR as a representative example to derive the semiparametrically efficient estimator. The corresponding results for the general causal measures are provided in the Supplementary Material to facilitate clarity and conciseness of the presentation. The estimators obtained in this section will serve as the foundation for the federated learning approach in $\S$\ref{fed}. 

Let $\psi^{(k)}_{1}=E\{Y(1)|S=k\}$ and $\psi^{(k)}_{0}=\mathbb E\{Y(0)|S=k\}$ for $k\in[[K]]$. Before introducing the EIF of $\psi^{(0)}$, we first focus on deriving the EIF of $\psi^{(0)}_{1}$ for the following reasons: (\romannumeral1) Under Condition $1^{(1:K)}$, this can be achieved without imposing restrictions on $\mu_0^{(0)}(x)$, while only restricting $\mu_1^{(0)}(x)$. Without restrictions, the EIF of $\psi^{(0)}_0$ is trivial; (\romannumeral2) Under Condition $2^{(1:K)}$, the derivation procedures for the EIFs of $\psi^{(0)}_{1}$ and $\psi^{(0)}_{0}$ are identical. 
 
 Using semiparametric theory, we derive efficient and robust estimators for $\psi^{(0)}_{1}$ under Condition $1^{(1:K)}$ and Condition $2^{(1:K)}$, which are presented in Theorem \ref{MR1} and \ref{MR2}, respectively.

\begin{theorem}
\label{MR1}
    Under Assumptions $1^{(0:K)}$, $2^{(0:K)}$ and \ref{inje}, as well as Condition $1^{(1:K)}$, we denote $\tau(X):=\tau^{(0)}(X)=\cdots=\tau^{(K)}(X)$, then the EIF of $\psi^{(0)}_{1}$ is
    $$\varphi_{1}^{(0)}(V)=\frac{I(S=0)\left\{\mu_1^{(0)}(X)-\psi^{(0)}_{1}\right\}+\sum_{k=0}^K\frac{C_{1k}}{\sum_{j=0}^{K}C_{1j}}H_1^{(k)}(V)}{\operatorname{pr}(S=0)}.$$
    
    Here,
    \begin{align*}
        H_1^{(k)}(V)&=\frac{I(S=k)q^{(k)}(X)\epsilon^{(k)}(X)}{\mu_{0}^{(k)}(X)}\left\{\frac{A\mu_{0}^{(0)}(X)}{\pi^{(k)}(X)}-\frac{(1-A)\mu_{1}^{(0)}(X)}{1-\pi^{(k)}(X)}\right\}\\
        &+\frac{I(S=0)(1-A)}{1-\pi^{(0)}(X)}\frac{\epsilon^{(0)}(X)\mu_{1}^{(k)}(X)}{\mu_{0}^{(k)}(X)} \quad \text{for } k\in[[K]],
    \end{align*}
    where $\epsilon^{(k)}(X)=Y-\mu_{0}^{(k)}(X)$.
    
    The ratio $C_{10}/C_{1k}=C_{10}(X)/C_{1k}(X)$ is
    \begin{align*}
    \frac{C_{10}}{C_{1k}}&=\frac{q^{(k)}(X)\pi^{(0)}(X)\sigma^2_{1,k}(X)}{\pi^{(k)}(X)\sigma^2_{1,0}(X)}\left\{\frac{\mu_{0}^{(0)}(X)}{\mu_{0}^{(k)}(X)}\right\}^{2}
    +\frac{q^{(k)}(X)\pi^{(0)}(X)\mu_{0}^{(0)}(X)\sigma^2_{0,k}(X)}{\{1-\pi^{(k)}(X)\}\sigma^2_{1,0}(X)}\left\{\frac{\mu_{0}^{(0)}(X)\tau(X)}{\mu_{0}^{(k)}(X)}\right\}^{2}\\
    &+\frac{\pi^{(0)}(X)\sigma^2_{0,0}(X)\{\tau(X)\}^{2}}{\{1-\pi^{(0)}(X)\}\sigma^2_{1,0}(X)}\sum_{l\neq 0}\frac{g^{(k)}(X)}{g^{(l)}(X)} \quad \text{for } k\in[[K]]\backslash\{0\},
\end{align*}
where $\sigma^2_{a,k}(X)=\operatorname{var}(Y|A=a,S=k,X)$ for $k\in[[K]],\ a=0,1$, and
$g^{(k)}(X)=[\sigma^2_{1,k}(X)/\pi^{(k)}(X)+\sigma^2_{0,k}(X)\{\tau(X)\}^2/\{1-\pi^{(k)}(X)\}]/[\operatorname{pr}(S=k|X)\{\mu_{0}^{(k)}(X)\}^{2}].$
\end{theorem}

\begin{remark}
    Notice that $H_1^{(0)}(V)=AI(S=0)\{Y-\mu_{1}^{(0)}(X)\}/\pi^{(0)}(X)$ remains the same no matter which causal measure we select. This EIF can also be written in another way: 
    $$\varphi_{1}^{(0)}(V)=\sum_{k=0}^K\frac{C_{1k}}{\sum_{j=0}^{K}C_{1j}}\frac{I(S=0)\left\{\mu_1^{(0)}(X)-\psi^{(0)}_{1}\right\}+H_1^{(k)}(V)}{\operatorname{pr}(S=0)},$$
    which can be seem as a weighted average of over $K+1$ augmented estimators with weights $C_{1k}/\sum_{j=0}^K C_{1j}$. These weights can be computed from the ratios $C_{10}/C_{1k}$ for $k\in[[K]]\backslash\{0\}$ alone and are constructed to ensure orthogonality of the EIF to nuisance directions. This construction enables efficient estimation even under heterogeneity across sites and forms the basis for semiparametric estimators that can incorporate flexible, possibly nonparametric nuisance function estimators while still achieving efficiency under suitable conditions.
\end{remark}

\begin{remark}
    We next examine the weights of each term. For each site-specific component, a smaller value of $C_{10}/C_{1k}$ increases the weight $C_{1k}/\sum_{j=0}^K C_{1j}$. Notice that a smaller variance of the treatment or control group outcomes in site $k$, corresponding to more precise outcome information, increases the weight assigned to that site. Similarly, a smaller $q^{(k)}(X)$ indicates that, given the covariates, individuals are more likely to appear in site $k$, which implies greater reliance on that site and therefore a larger weight. Moreover, a larger $\mu_0^{(k)}(X)$ also contributes to a larger weight, since for a fixed $\tau(X)$, an increase in $\mu_0^{(k)}(X)$ reduces the variability of $\mu_1^{(k)}(X)$, leading to more stable estimation.
\end{remark}

\begin{theorem}
\label{MR2}
    Under Assumptions $1^{(0:K)}$, $2^{(0:K)}$ and \ref{inje}, as well as Condition $2^{(1:K)}$, we denote $\mu_a(X):=\mu_a^{(0)}(X)=\cdots=\mu_a^{(K)}(X)$ for $a=0,1$, then the EIF of $\psi^{(0)}_{1}$ is
    $$\varphi_{1}^{(0)}(V)=\frac{I(S=0)\left\{\mu_1^{(0)}(X)-\psi^{(0)}_{1}\right\}+\sum_{k=0}^K\frac{C_{2k}}{\sum_{j=0}^{K}C_{2j}}H_2^{(k)}(V)}{\operatorname{pr}(S=0)},$$
    where $H_2^{(k)}(V)=I(S=k)Aq^{(k)}(X)\{Y-\mu_{1}(X)\}/\pi^{(k)}(X)$. The ratio $C_{20}/C_{2k}=C_{20}(X)/C_{2k}(X)=q^{(k)}(X)\pi^{(0)}(X)\sigma^2_{1,k}(X)/\{\pi^{(k)}(X)\sigma^2_{1,0}(X)\}$ where $\sigma^2_{a,k}(X)=\operatorname{var}(Y|A=a,S=k,X)$ for $k\in[[K]],\ a=0,1$.
\end{theorem}

\begin{remark}
    From Theorem \ref{MR2}, the EIF for $\psi^{(0)}_1$ remains the same across causal measures satisfying Assumption \ref{inje}. However, as shown in Theorem \ref{MR1}, the EIF varies with the causal measure. This distinction is natural: under Condition $2^{(1:K)}$, separate restrictions are imposed on the treated and control group, whereas under Condition $1^{(1:K)}$, the two groups are linked through the restriction on $\tau{(X)}$ function.
\end{remark}

\begin{remark}
    The EIF under Condition $2^{(1:K)}$ has a similar form to that under Condition $1^{(1:K)}$, differing only in the definitions of $H_1^{(k)}(V)$ and $H_2^{(k)}(V)$. As in Theorem \ref{MR1}, the EIF in Theorem \ref{MR2} can be expressed as a weighted average with weights $C_{2k}/\sum_{j=0}^KC_{2j}$. A smaller value of $C_{20}/C_{2k}$ corresponds to a larger weight $C_{2k}/\sum_{j=0}^KC_{2j}$. Similarly, a smaller $q^{(k)}(X)$ increases the weight. Notably, $C_{20}/C_{2k}$ depends only on the conditional variance of the treated group in site $k$ and is independent of the control group variance, and unlike the ratio in Theorem \ref{MR1}, the weight does not depend on $\mu_0^{(k)}(X)$. These differences arise because Condition \ref{con2} imposes separate assumptions for treated and control groups, whereas Condition \ref{con1} links them via transportability of the treatment effect.
\end{remark}

While both Condition \ref{con1} and Condition \ref{con2} allow identification of the target causal estimand, Condition \ref{con2} imposes stricter structure on the outcome models. It is natural to expect that the resulting semiparametric bound of the EIF in Theorem \ref{MR2} has more efficiency gain when Condition $2^{(1:K)}$ is met. The following theorem formalizes this intuition.
\begin{theorem}
\label{semi order}
    When Condition $2^{(1:K)}$ holds, the semiparametric efficiency bound of the EIF for either $\psi_1^{(0)}$ or $\psi^{(0)}$ under Condition $2^{(1:K)}$ is smaller than that under Condition $1^{(1:K)}$.
\end{theorem}

\begin{remark}
This result implies that, imposing stricter conditions can improve the semiparametric bound, as the estimator leverages the additional structure to reduce uncertainty. Our simulation results in $\S$\ref{results} also illustrate this phenomenon. 
\end{remark}

For EIF from Theorem \ref{MR1}, the estimator of $\psi_1^{(0)}$ can be obtained by solving the function 
$\mathbb{P}_n(\varphi_{1}^{(0)}(V;\psi_1^{(0)},\{\widehat\pi^{(k)},\widehat\mu_0^{(k)}\}_{k=0}^K,\{\widehat q^{(k)}\}_{k=1}^K,\widehat\tau))=0$, where $\varphi_1^{(0)}(\cdot)$ denotes the full EIF depending on the the collection of nuisance functions $\{\pi^{(k)},\mu_0^{(k)}\}_{k=0}^K$, $\{q^{(k)}\}_{k=1}^K$ and $\tau$. $\{\widehat\pi^{(k)},\widehat\mu_0^{(k)}\}_{k=0}^K$, $\{\widehat q^{(k)}\}_{k=1}^K$ and $\widehat\tau$ are their estimators. Here, $\mathbb P_n$ represents the empirical measure, i.e., $\mathbb P_n(h(V))=\sum_{i=1}^nh(V_i)$ for arbitrary function $h$. Similarly, for EIF from Theorem \ref{MR2}, the principled estimator is obtained by solving the function 
$\mathbb{P}_n(\varphi_{1}^{(0)}(V;\psi_1^{(0)},\{\widehat\pi^{(k)}\}_{k=0}^K,\{\widehat q^{(k)}\}_{k=1}^K,\widehat\mu_1))=0$. 

Similarly, as previously mentioned, we can obtain the estimator $\widehat\psi^{(0)}_{0}$ based on its EIF $\varphi_0^{(0)}$. The EIF for $\psi^{(0)}$, denoted $\varphi^{(0)}$, can be derived using the pathwise derivative. For instance, for RR, it can be calculated based on Lemma S2 in \citet{jiang2022multiply}. The corresponding estimator for $\psi^{(0)}$ can be calculated as $\widehat\psi^{(0)}=m(\widehat\psi^{(0)}_{0},\widehat\psi^{(0)}_{1})$. 

\subsection{Theoretical properties}
\label{property}
In this section, we use the RR as our causal measure example because we showed an explicit analytical form for its EIF. The properties of other causal measures can be derived in a similar manner. In this section, we demonstrate that our EIF-based estimator possesses the following properties: (\romannumeral1) multiple robustness: $\widehat\psi^{(0)}_1$ remains consistent for $\psi_1^{(0)}$ even when some of the nuisance functions are misspecified simultaneously; (\romannumeral2) rate robustness: $\widehat\psi^{(0)}$ remains the parametric-rate consistent for $\psi^{(0)}$ after incorporating flexible methods for estimating the nuisance functions under some regularity conditions; (\romannumeral3) local efficiency: $\widehat\psi^{(0)}$ achieves the semiparametric efficiency bound if the nuisance functions are correctly specified or the nuisance function estimation satisfy the same regularity conditions.

The following theorems demonstrate the multiple robustness of the estimators constructed from the EIFs under Condition $1^{(1:K)}$ and Condition $2^{(1:K)}$. Here, we focus on the multiple robustness of $\widehat\psi^{(0)}_1$. For the EIF in Theorem \ref{MR1}, the relevant nuisance functions are $\pi^{(k)}(x)$ and $\mu_0^{(k)}(x)$ for $k\in[[K]]$, together with $\tau(x)$ and $q^{(k)}(x)$ for $k\in[[K]]\backslash{0}$. For the EIF in Theorem \ref{MR2}, they are $\pi^{(k)}(x)$ for $k\in[[K]]$, $\mu_1(x)$, and $q^{(k)}(x)$ for $k\in[[K]]\backslash{0}$.

\begin{theorem}
    Each estimator in Theorem \ref{MR1} is multiply robust in the sense that it is consistent for $\psi^{(0)}_1$ under (\romannumeral1) $\tau(x)$ is correctly specified and  no $\pi^{(k)}(x)$ and $\mu_0^{(k)}(x)$ are simultaneously misspecified for $k\in[[K]]$; (\romannumeral2) Only $\tau(x)$ is misspecified.
\end{theorem}

\begin{theorem}
    Each estimator in Theorem \ref{MR2} is multiply robust in the sense that it is consistent for $\psi^{(0)}_1$ under (\romannumeral1) $\mu_1(x)$ is correctly specified, and all the other nuisancne functions can be misspecified simultaneously; (\romannumeral2) Only $\mu_1(x)$ is misspecified.
\end{theorem}

\begin{remark}
    When the estimator is constructed under conditions assuming that the conditional causal measure is identical across sites, $\tau(x)$ serves as the linking function, such that its misspecification precludes misspecification of the remaining nuisance functions. Conversely, under conditions assuming that the conditional outcome is the same across sites, $\mu_1(X)$ plays this role, and its misspecification likewise precludes misspecification of the other nuisance functions.
\end{remark}

Multiply robust estimators were initially developed to provide robustness against parametric misspecification, but they are now known to also be robust to approximation errors using machine learning methods. 
We will investigate this new multiply robust feature and local efficiency property for the estimator $\widehat\psi^{(0)}$ under the two conditions separately, employing flexible semiparametric or nonparametric methods to estimate all nuisance functions in the estimators. Using flexible methods alleviates bias from the misspecification of parametric models. The following regularity conditions are required for the nuisance function estimators.

\begin{assumption}
\label{con1regul}
    For a random variable $Z$, define its $L_2$-norm as $\|Z\|=\{\mathbb E(Z^2)\}^{1/2}$. Assume (\romannumeral1) $\{\widehat\pi^{(k)},\widehat\mu_0^{(k)},\widehat q^{(k)},\widehat\tau\}\xrightarrow{p} \{\pi^{(k)},\mu_0^{(k)},q^{(k)},\tau\}$ for $k\in[[K]]$; (\romannumeral2) $\|\widehat{\pi}^{(k)}(X)-\pi^{(k)}(X)\|\times\|\widehat{\mu}_{0}^{(k)}(X)-\mu_{0}^{(k)}(X)\|=o_p(n^{-1/2})$ for $k\in[[K]]$; (\romannumeral3) $\|\widehat{\pi}^{(k)}(X)-\pi^{(k)}(X)\|\times\|\widehat{\tau}(X)-\tau(X)\|=o_p(n^{-1/2})$ for $k\in[[K]]$; (\romannumeral4) $\|\widehat{\mu}_{0}^{(k)}(X)-\mu_{0}^{(k)}(X)\|\times\|\widehat{\tau}(X)-\tau(X)\|=o_p(n^{-1/2})$ for $k\in[[K]]$; (\romannumeral5) $\|\widehat{q}^{(k)}(X)-q^{(k)}(X)\|\times\|\widehat{\tau}(X)-\tau(X)\|=o_p(n^{-1/2})$ for $k\neq 0$; (\romannumeral6) additional regularity conditions in the Supplementary Material.
\end{assumption}

\begin{assumption}
\label{con2regul}
    Assume (\romannumeral1) $\{\widehat\pi^{(k)},\widehat q^{(k)},\widehat\mu_1\}\xrightarrow{p} \{\pi^{(k)},q^{(k)},\mu_1\}$ for $k\in[[K]]$; (\romannumeral2) $\|\widehat{\pi}^{(k)}(X)-\pi^{(k)}(X)\|\times\{\|\widehat{\mu}_1(X)-\mu_1(X)\|\vee\|\widehat{\mu}_0(X)-\mu_0(X)\|\}=o_p(n^{-1/2})$ for $k\in[[K]]$; (\romannumeral3) $\|\widehat{q}^{(k)}(X)-q^{(k)}(X)\|\times\{\|\widehat{\mu}_1(X)-\mu_1(X)\|\vee\|\widehat{\mu}_0(X)-\mu_0(X)\|\}=o_p(n^{-1/2})$ $k\neq 0$; (\romannumeral4) additional regularity conditions in the Supplementary Material.
\end{assumption}

\begin{remark}
    Assumptions \ref{con1regul} and \ref{con2regul} are standard regularity conditions used in M-estimation to achieve rate robustness \citep{van2000asymptotic}. Under Condition $1^{(1:K)}$ with Assumption \ref{con1regul}, or under Condition $2^{(1:K)}$ with Assumption \ref{con2regul}, the proposed estimators in $\S$\ref{est} can incorporate flexible methods for estimating the nuisance functions while still achieving parametric-rate consistency.
\end{remark}

\begin{theorem}
\label{loceff}
Under Assumptions $1^{(0:K)}$ and $2^{(0:K)}$, the estimator $\widehat\psi^{(0)}$ is asymptotically normal if either: (\romannumeral1) Condition $1^{(1:K)}$ holds, and Assumption \ref{con1regul} is satisfied or all nuisance functions are correctly specified; or (\romannumeral2) Condition $2^{(1:K)}$ holds, and Assumption \ref{con2regul} is satisfied or all nuisance functions are correctly specified.

In either case, we have
$n^{1/2}(\widehat\psi^{(0)} - \psi^{(0)}) \xrightarrow{d} N(0, \mathbb{V}_{\psi^{(0)}})$, where
$\mathbb{V}_{\psi^{(0)}} = \mathbb{E}[\{\varphi^{(0)}(V)\}^2]$. Here, $\varphi^{(0)}(V)=\varphi^{(0)}(V;\{\pi^{(k)},\mu_0^{(k)}\}_{k=0}^K,\{ q^{(k)}\}_{k=1}^K,\tau)$ for case (\romannumeral1), and $\varphi^{(0)}(V)=\varphi^{(0)}(V;\{\pi^{(k)}\}_{k=0}^K,\{q^{(k)}\}_{k=1}^K,\mu_1,\mu_0)$ for case (\romannumeral2).
\end{theorem}

Theorem \ref{loceff} motivates variance estimation by $\widehat{\mathbb V}_{\psi^{(0)}}=n^{-1}\sum_{i=1}^n[\varphi^{(0)}(V_i;\{\widehat\pi^{(k)},\widehat\mu_0^{(k)}\}_{k=0}^K,\{\widehat q^{(k)}\}_{k=1}^K,\widehat\tau)]^2$ for case (\romannumeral1), and $\widehat{\mathbb V}_{\psi^{(0)}}=n^{-1}\sum_{i=1}^n[\varphi^{(0)}(V_i;\{\widehat\pi^{(k)}\}_{k=0}^K,\{\widehat q^{(k)}\}_{k=1}^K,\widehat\mu_1,\widehat\mu_0)]^2$ for case (\romannumeral2). Both are consistent estimators of $\mathbb V_{\psi^{(0)}}$ as established in Theorem \ref{loceff}.

\section{Heterogeneity-aware federated estimator}
\label{fed}
Throughout this section, the framework of our algorithm is based on the estimator derived using Theorem \ref{MR1}, as the corresponding EIF estimator tends to be more general and robust. Moreover, as demonstrated in the simulation studies in $\S$\ref{sim}, it exhibits superior performance across a wider range of scenarios. 

In practical setting, it is often unreasonable to assume that Assumptions \ref{unconf} and \ref{posi}, as well as Condition \ref{con1} hold across all source sites. In such cases, some source sites may provide relevant information for estimating the estimand, while other sites may not. In this section, we present an approach that combines information from both target site and source sites in a data-adaptive manner. We only assume that the identification assumptions hold for the target site, namely Assumptions $1^{(0)}$ and $2^{(0)}$. The algorithm is designed to detect source sites within which unconfoundedness, positivity assumptions, or Condition \ref{con1} may not hold, and to exclude these sites accordingly. This enables valid and efficient estimation, even in the presence of assumption violations and heterogeneity across sites. Our approach consists of two steps. In the first step, following \citet{liu2024multi}, we assign a weight for each site, which motivates the initial weighted estimator $\widehat\psi_{\text{fw}}^{(0)}$. In the second step, we propose a selection procedure based on the weights obtained in the first step. The result of the selection determines our final estimator, $\widehat\psi_{\text{fs}}^{(0)}$.  

We first define a new estimator $\widehat\psi_1^{(0)<k>}$ for $\psi_1^{(0)}$, obtained by applying Theorem \ref{MR1} using only data from the target site and source site $k$. In particular, $\widehat\psi_1^{(0)<0>}$ is defined as the standard Augmented Inverse Probability Weighting (AIPW) estimator based solely on target-site data. The corresponding EIFs are denoted by $\varphi_1^{(0)<k>}$. Analogous superscripts are used for estimators and EIFs associated with $\psi_0^{(0)}$ and $\psi^{(0)}$. To aggregate information from the target and source sites, we first compute the discrepancy measures $\widehat\delta^{(k)}=|\widehat\psi^{(0)<k>}_1-\widehat\psi^{(0)<0>}_1|$. We then solve for federated weights $\widehat{\pmb w}=(\widehat w_0,\widehat w_1,\cdots,\widehat w_{K})$ that minimize the following loss:
$$
    Q(\pmb w)=\mathbb{P}_{n}\left\{\left(\widehat\varphi_1^{(0)<0>}-\sum_{k\neq 0}w_{k}\widehat\varphi_1^{(0)<k>}\right)^{2}\right\}+\lambda_n\sum_{k\neq 0}|w_k|\left(\widehat\delta^{(k)}\right)^2,
$$subject to $0\leq w_k\leq1$ for $k\in[[K]]$, and $\sum_{k=0}^Kw_k=1$. Here, $\widehat\varphi_1^{(0)<0>}=\varphi^{(0)<0>}_1(V;\widehat{\psi}^{(0)<0>}_1,\widehat{\pi}^{(0)},\widehat{\mu}_1^{(0)})$ and  $\widehat\varphi_1^{(0)<k>}=\varphi^{(0)<k>}_1(V;\widehat{\psi}^{(0)<0>}_1, \{\widehat\pi^{(l)},\widehat\mu_0^{(l)}\}_{l=0,k},\widehat q^{(k)},\widehat\tau)$. $\lambda_n$ is a tuning parameter chosen by cross-validation. One federated weighted estimator is given by:
$$\widehat\psi^{(0)}_{\text{fw}}=m\left(\widehat\psi_0^{(0)<0>},\sum_{k=0}^K\widehat w_{k}\widehat\psi^{(0)<k>}_1\right).$$

Denote $\mathcal{S}=\left\{k\in[[K]]:\tau^{(k)}(X)=\tau^{(0)}(X)\right\}$, which represents the set of sites meeting Condition \ref{con1}. The penalization procedure has the following property:

\begin{theorem}
\label{fw}
    Assume that Assumptions $1^{(0)}$, $2^{(0)}$, \ref{inje} and \ref{con1regul} hold. Suppose $n^{-1/2}\lambda_n\rightarrow0$ and $n^{(\gamma-1)/2}\lambda_n\rightarrow\infty$ for $\gamma>0$. We have $\widehat w_k\xrightarrow{p} 0$, for $k\notin \mathcal S$.
\end{theorem}

From Theorem \ref{fw}, we see that as $n\rightarrow\infty$, the weights of sites not in $\mathcal S$ shrink to $0$ in probability. However, in practical settings with finite sample sizes, these weights may not explicitly reach $0$. Therefore, we proceed with a selection procedure to exclude sites outside $\mathcal S$ whose weights remain nonzero owing to finite-sample effects. This selection procedure serves as a second safeguard in site selection, reinforcing the selection consistency.

We want to select a threshold $e>0$ to choose all the source sites meeting Condition \ref{con1}, that is, those with $\widehat w_k\geq e$ to apply the EIF-based estimator derived from Theorem \ref{MR1}. We denote the final estimator for $\psi^{(0)}$ after selection with threshold $e$ as $\widehat\psi^{(0)}_e$ for $0<e\leq1$. The choice of $e$ balances bias and the variance: a larger $e$ retains fewer source sites, reducing bias but increasing variance, whereas a smaller $e$ includes more source sites, reducing variance but increasing bias. Thus, we use mean squared error (MSE) to guide the selection of $e$ and propose a data-adaptive procedure to directly minimize the MSE of $\widehat\psi^{(0)}_e$.

We decompose $\operatorname{MSE}(e)=\mathbb E\{(\widehat\psi^{(0)}_e-\psi^{(0)})^2\}=\{\mathbb E(\widehat\psi^{(0)}_e)-\psi^{(0)}\}^2+\mathbb V(\widehat\psi^{(0)}_e)$. Since $\widehat\psi^{(0)<0>}$ is consistent for $\psi^{(0)}$, we approximate $\{\mathbb E(\widehat\psi^{(0)}_e)-\psi^{(0)}\}^2$ by $\{\mathbb E(\widehat\psi^{(0)}_e)-\widehat\psi^{(0)<0>}\}^2=\mathbb E\{(\widehat\psi^{(0)}_e-\widehat\psi^{(0)<0>})^2\}-\mathbb V(\widehat\psi^{(0)}_e-\widehat\psi^{(0)<0>})$. We then use $(\widehat\psi^{(0)}_e-\widehat\psi^{(0)<0>})^2$ to estimate $\mathbb E\{(\widehat\psi^{(0)}_e-\widehat\psi^{(0)<0>})^2\}$ and apply bootstrap to estimate $\operatorname{cov}(\widehat\psi^{(0)}_e,\widehat\psi^{(0)<0>})$ and $\mathbb V(\widehat\psi^{(0)<0>})$. Then we can approximate $\operatorname{MSE}(e)$ by
$$\widehat{\operatorname{MSE}}(e)=(\widehat\psi^{(0)}_e-\widehat\psi^{(0)<0>})^2+2\widehat{\operatorname{cov}}(\widehat\psi^{(0)}_e,\widehat\psi^{(0)<0>})-\widehat{\mathbb V}(\widehat\psi^{(0)<0>}).$$

After this, the choice of $e$ follows the steps below:

\begin{algorithm}
\caption{Adaptive selection threshold.}
\label{alg.threshold}
\begin{algorithmic}[1]
\REQUIRE Threshold grid $E$, number of bootstrap samples $B$
\FOR{$e \in E$}
    \FOR{$b=1,2,\cdots,B$}
        \STATE Calculate $\widehat \psi_{e}^{(0)[b]}$ and $\widehat\psi^{(0)<0>[b]}$ from the $b$-th bootstrap sample.
    \ENDFOR
    \STATE Compute $\widehat{\operatorname{cov}}(\widehat \psi_{e}^{(0)},\widehat\psi^{(0)<0>})=(B-1)^{-1}\sum_{b=1}^B(\widehat \psi_{e}^{(0)[b]}-\overline{\psi}^{(0)}_e)(\widehat\psi^{(0)<0>[b]}-\overline\psi^{(0)<0>})$,\\
    where $\overline{\psi}^{(0)}_e=B^{-1}\sum_{b=1}^B\widehat \psi_{e}^{(0)[b]}$ and $\overline\psi^{(0)<0>}=B^{-1}\sum_{b=1}^B\widehat\psi^{(0)<0>[b]}$.
    \STATE Compute $\widehat{\mathbb V}(\widehat\psi^{(0)<0>})=(B-1)^{-1}\sum_{b=1}^B\left(\widehat\psi^{(0)<0>[b]}-\overline{\psi}^{(0)<0>}\right)^2$.
    \STATE Compute $\widehat{\operatorname{MSE}}(e)=\left(\widehat \psi_{e}^{(0)}-\widehat\psi^{(0)<0>}\right)^2+2\widehat{\operatorname{cov}}(\widehat \psi_{e}^{(0)},\widehat\psi^{(0)<0>})-\widehat{\mathbb V}(\widehat\psi^{(0)<0>})$.
\ENDFOR
\ENSURE $e^*=\max\{\arg\min_{e\in E}\widehat{\operatorname{MSE}}(e)\}$
\end{algorithmic}
\end{algorithm}

\begin{remark}
    Due to the consistency of bootstrap, the variances of both $\widehat\psi_{e^*}^{(0)}$ and $\widehat\psi^{(0)<0>}$ are naturally obtained in the process. Additionally, since $\widehat\psi^{(0)}_e$ may remain the same for different values of $e$ (as they select the same set of sites), we only need to choose one of the values in $\arg\min_{e\in E}\widehat{\operatorname{MSE}}(e)$ and some iterations can be skipped to reduce computation time. 
\end{remark}

The final estimator is $\widehat\psi_{\text{fs}}^{(0)}=\widehat\psi_{e^*}^{(0)}$. The asymptotic properties for $\widehat\psi_{\text{fs}}^{(0)}$ follow directly from applying Theorem \ref{loceff} to the data from $\widetilde{\mathcal S}_{e^*}$, where $\widetilde{\mathcal{S}}_{e^*}$ denotes the sites selected by threshold $e^*$. This is formalized in the following theorem.
\begin{theorem}
    Under Assumptions $1^{(0)}$, $2^{(0)}$, \ref{inje} and \ref{con1regul}, we have $n^{1/2}(\widehat\psi^{(0)}_{\textnormal{fs}}-\psi^{(0)})\xrightarrow{d}N(0,\mathbb V_{\psi^{(0)}})$ where $\mathbb V_{\psi^{(0)}}=\mathbb E\{[\varphi^{(0)}(V;\{\pi^{(k)},\mu_0^{(k)}\}_{k\in\mathcal S},\{q^{(k)}\}_{k\in\mathcal S\backslash\{0\}},\tau)]^2\}$.
\end{theorem}

\begin{remark}
    Note that both the federated selection method and its asymptotical results can be applied to the estimators obtained according to Theorem \ref{MR1} or \ref{MR2} for any causal measure meeting Assumption \ref{inje}. That is, this federated selection method is a general method for the EIF estimators across multiple sites. 
\end{remark}

\section{Simulation studies}
\label{sim}
\subsection{Overview of simulation studies}
We evaluate our proposed methods using Monte Carlo simulations under two main scenarios with respective sub-cases. In $\S$\ref{results}, we evaluate the performance of estimators across the two scenarios based on the following criteria: (\romannumeral1) bias between the point estimate and the true value; (\romannumeral2) standard error of estimate; (\romannumeral3) MSE; (\romannumeral4) coverage probability of the confidence interval (CI).

We conduct $M=500$ replications with total sample size $n=1000$ across three sites, $k\in\{0,1,2\}$, with $n=\sum_{k=0}^2n_k$. Full data-generating details are provided in the Supplementary Material. Define $b^{(k)}_\mu(x)=\mu^{(k)}_0(x)-\mu^{(0)}_0(x)$ and $b^{(k)}_\tau(x)=\tau^{(k)}(x)-\tau^{(0)}(x)$ for $k\in[[K]]$. We conduct the simulation study under the following framework:
\begin{enumerate}[label=Scenario \arabic*:, align=left, labelwidth=2cm, leftmargin=2cm]
    \item Treatment effect is transportable (Condition $1^{(1:2)}$ holds), i.e., $b^{(1)}_\tau(x)=b^{(2)}_\tau(x)=0$.
        \begin{enumerate}[label=Case 1.\arabic*:]
        \item No potential outcome shift: $b^{(1)}_{\mu}(x)=b^{(2)}_\mu(x)=0$.
        \item Presence of potential outcome shift: $b^{(1)}_\mu(x)=-10,\ b^{(2)}_\mu(x)=15$.
        \end{enumerate}
    
    \item Treatment effect may not be transportable (Conditions $1^{(1:2)}$ may fail), with outcome shifts $b^{(1)}_\mu(x)=-10,\ b^{(2)}_\mu(x)=15$.
        \begin{enumerate}[label=Case 2.\arabic*:]
        \item All sites transportable: $b^{(1)}_\tau(x)=\ b^{(2)}_\tau(x)=0$.
        \item Some sites transportable: $b^{(1)}_\tau(x)=0,\ b^{(2)}_\tau(x)=5$.
        \item No sites transportable: $b^{(1)}_\tau(x)=5,\ b^{(2)}_\tau(x)=5$.
        \end{enumerate}
\end{enumerate}

Scenario 1 investigates multiple robustness and efficiency of EIF-based estimators under full transportability, where the federated estimator is not relevant. Scenario 2 examines the federated estimator under varying transportability across sites.

For notational convenience, the EIF-based estimators derived using Theorem \ref{MR1} and Theorem \ref{MR2} are referred to as MR1 and MR2, respectively. MR refers to the multiple robustness property of these estimators, analogous to the double robustness (DR) of the standard AIPW estimator. Competing methods are summarized in Table \ref{estimators}. Scenario 1 compares MR1, MR2 and DR-t under various nuisance function misspecifications. The specific types of misspecification are detailed in the $\S$\ref{results}. Scenario 2 compares MR1, DR-t, FWMR1 and FSMR1.

\begin{table}[h]
    \centering
    \caption{Competing methods.}
    \begin{tabular}{ll}
    Estimators & Details\\
    MR1 & the EIF estimator obtained under Condition 1, which is multiply robust\\
    MR2 & the EIF estimator obtained under Condition 2, which is multiply robust\\
    DR-t & the AIPW estimator using the target site data only, which is doubly robust\\
    FWMR1 & the federated weighted estimator based on MR1\\
    FSMR1 & the federated selective estimator based on MR1
    \end{tabular}
    \label{estimators}
\end{table}

\subsection{Results}
\label{results}
In Scenario 1, we compare the proposed MR1 estimator against MR2 and DR-t under two sub-cases. To demonstrate multiple robustness properties, we evaluate MR1 and MR2 under four representative misspecification types based on Assumption \ref{con1regul} and \ref{con2regul}: (\romannumeral1) No nuisance function misspecified; (\romannumeral2) $\pi^{(1)}$, $\pi^{(2)}$, $\pi^{(3)}$, $q^{(2)}$ and $q^{(3)}$ are misspecified; (\romannumeral3) $\pi^{(3)}$, $\mu_0^{(1)}$, $\mu_0^{(2)}$, $q^{(2)}$ and $q^{(3)}$ are misspecified; (\romannumeral4) $\tau$ is misspecified. Since only $\pi^{(1)}$, $\mu_0^{(1)}$ and $\tau$ are the nuisance functions for the DR-t estimator, we define its misspecification types by taking the intersection of these three function with the misspecified functions in each of (\romannumeral1)-(\romannumeral4).

\begin{figure}[htbp]
	\centering
	\begin{subfigure}{0.49\linewidth}
		\centering
		\includegraphics[width=0.95\linewidth]{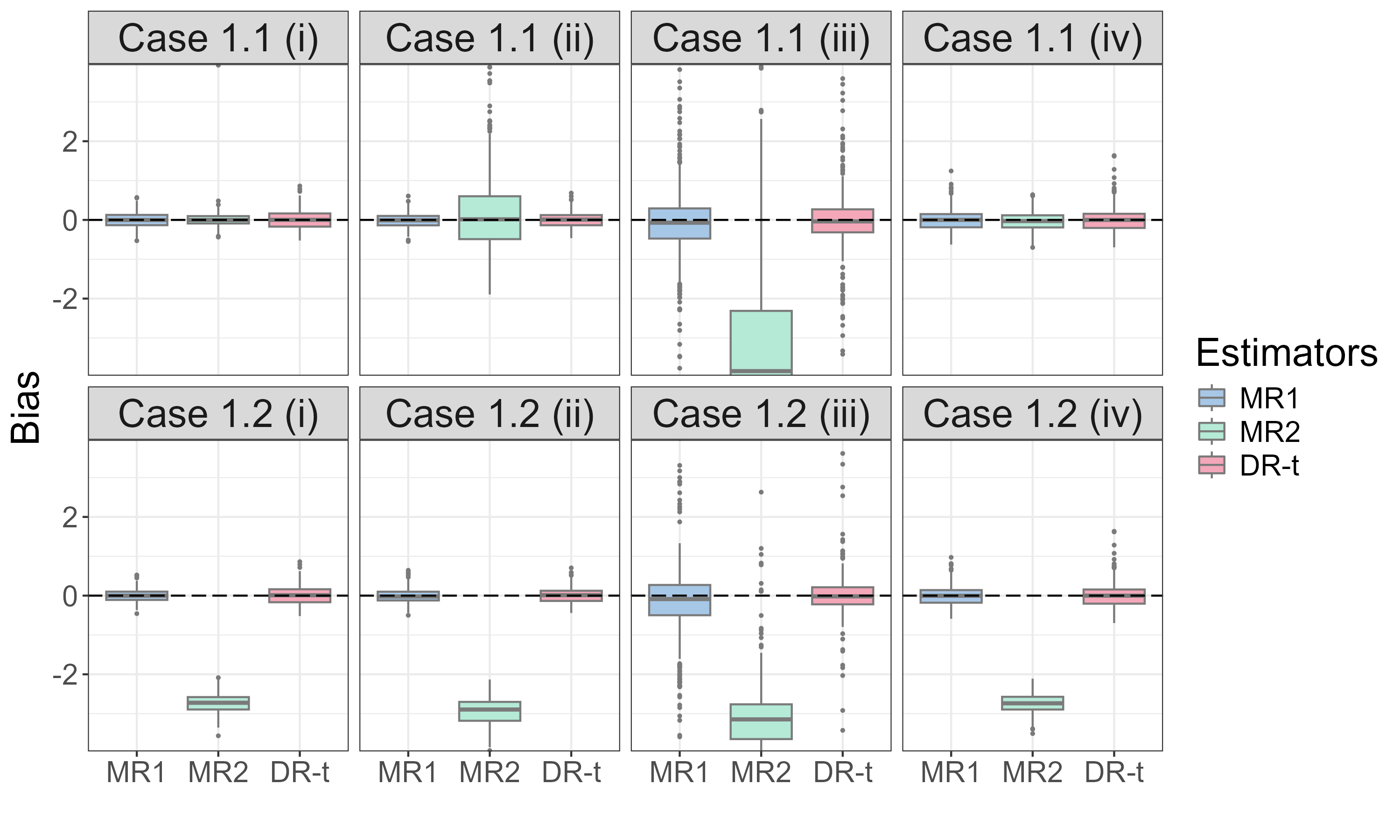}
		\caption{Bias}
		\label{sim1}
	\end{subfigure}
	\begin{subfigure}{0.49\linewidth}
		\centering
		\includegraphics[width=0.95\linewidth]{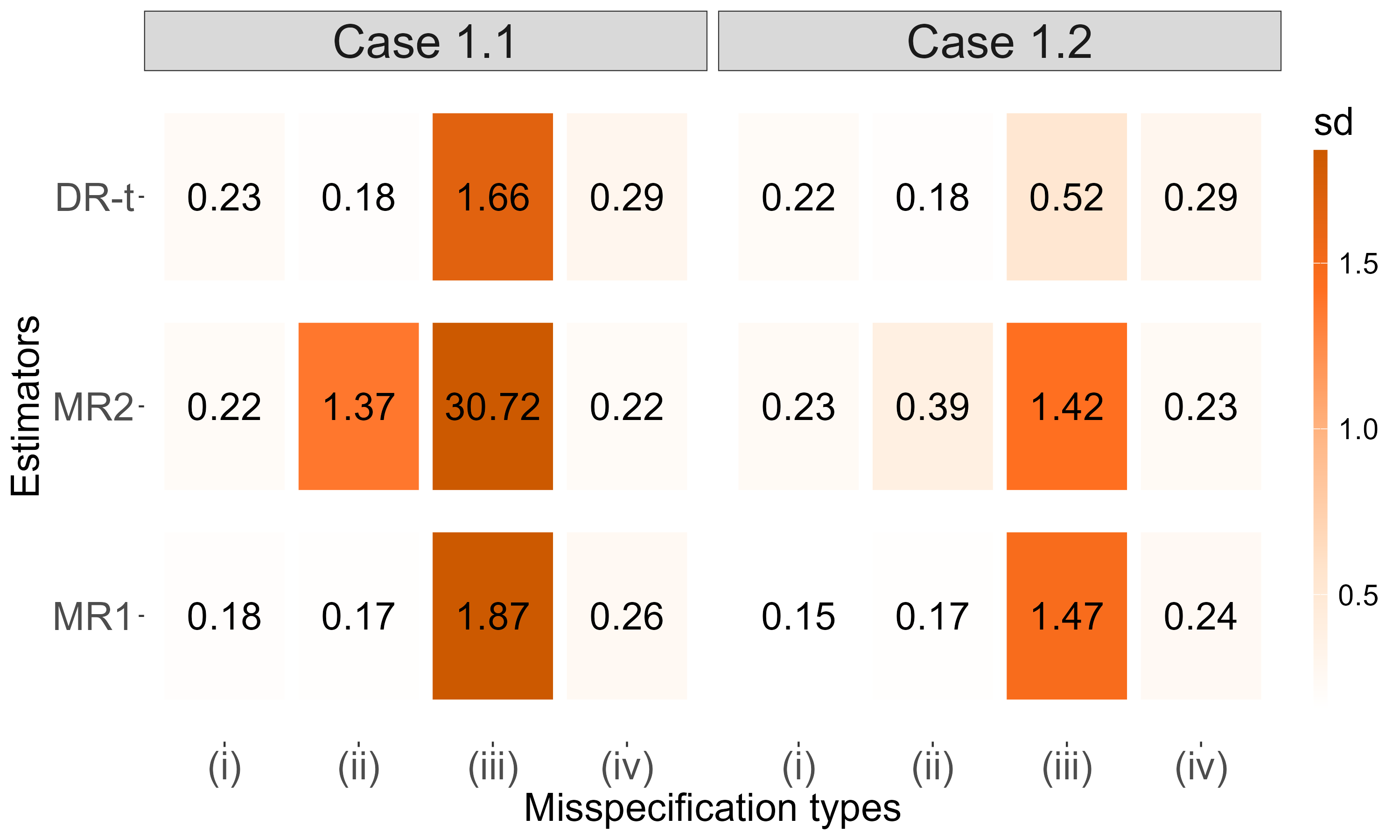}
		\caption{Standard error}
		\label{sim1_sd}
	\end{subfigure}

	\vspace{8mm}
    
	\begin{subfigure}{0.49\linewidth}
		\centering
		\includegraphics[width=\linewidth]{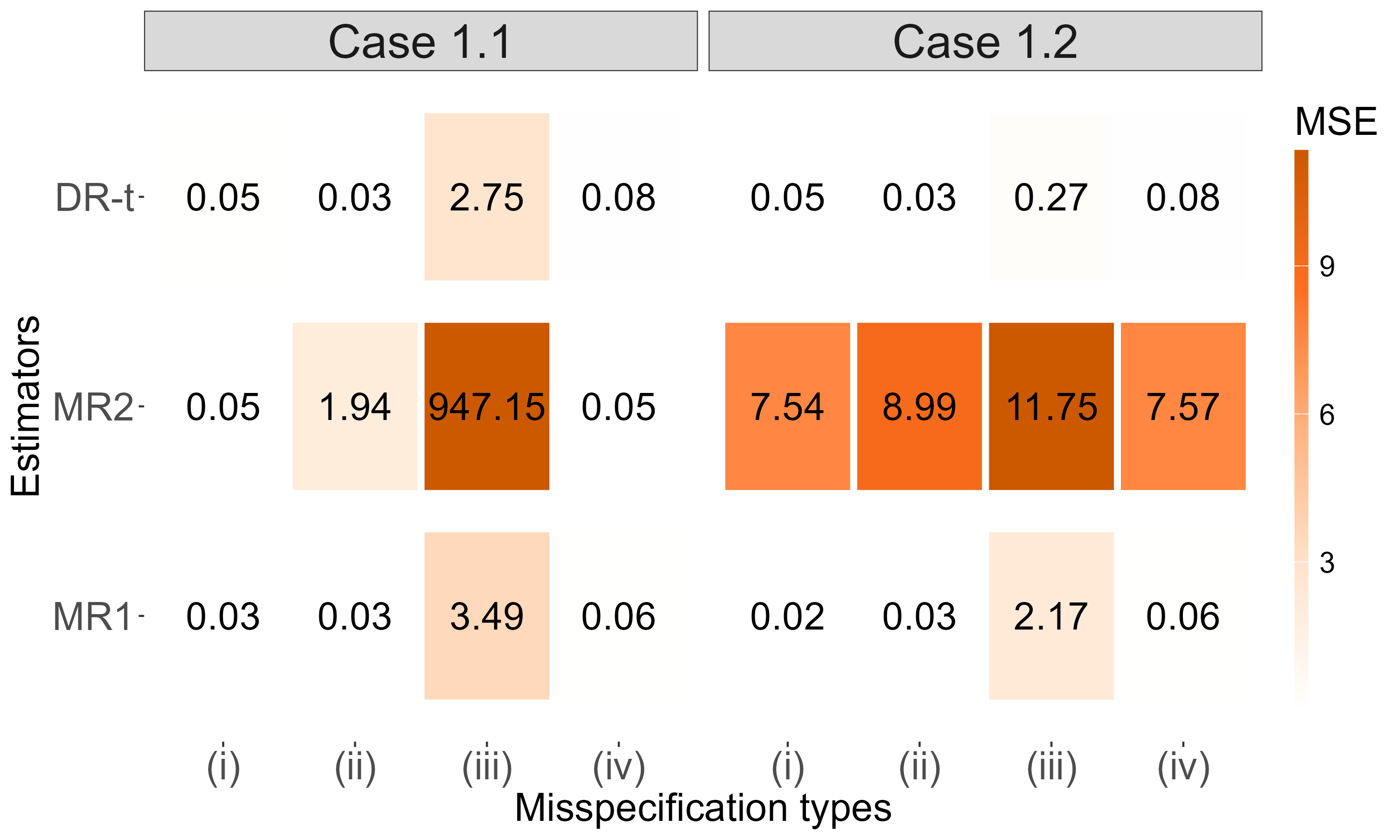}
		\caption{Mean squared error}
		\label{sim1_mse}
	\end{subfigure}
	\begin{subfigure}{0.49\linewidth}
		\centering
		\includegraphics[width=\linewidth]{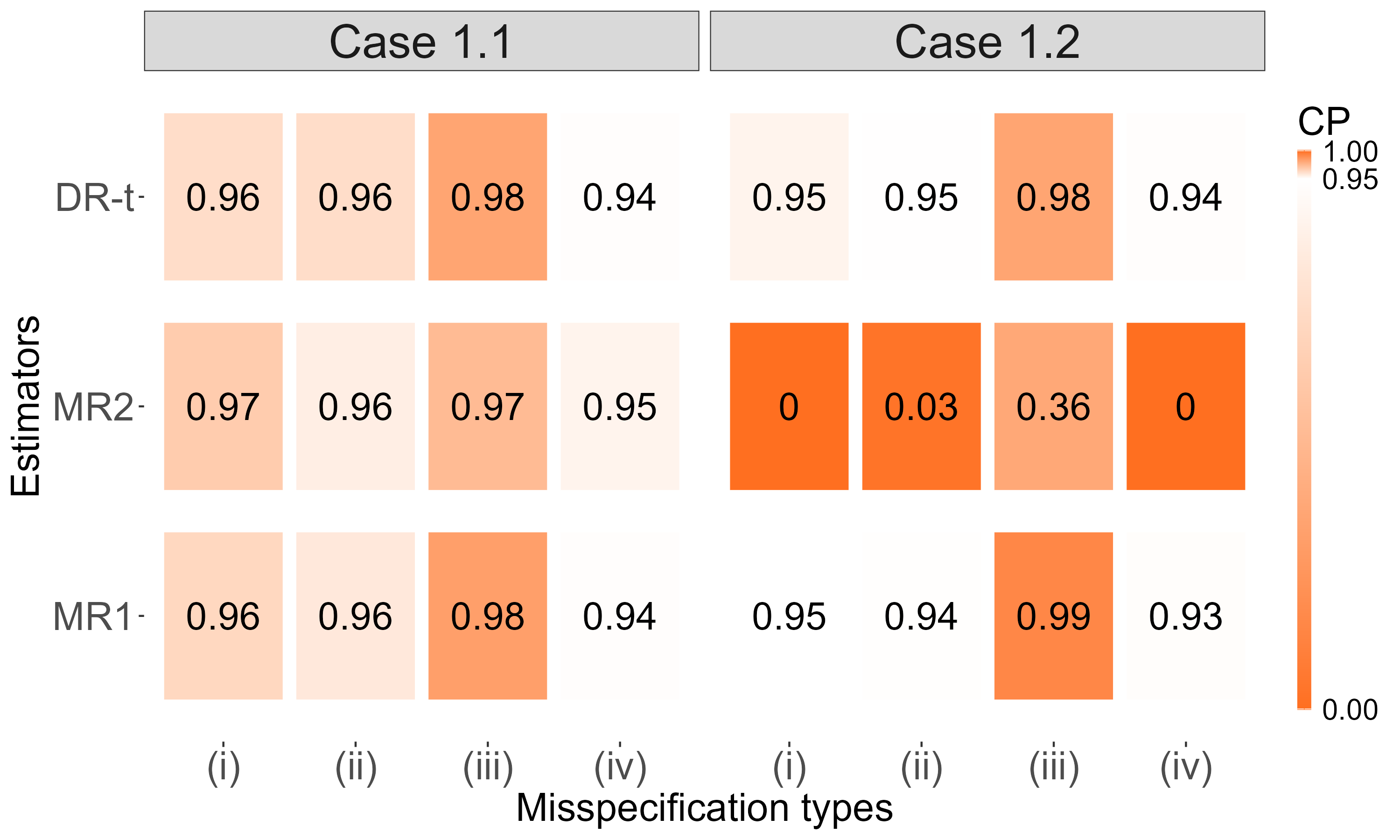}
		\caption{Coverage probability}
		\label{sim1_cp}
	\end{subfigure}
    \caption{Simulation results of Scenario 1.}
\end{figure}

Figure \ref{sim1} presents the bias of point estimates under the two potential outcome shift cases, across the listed misspecification types. When Condition $2^{(1:K)}$ holds, the MR2 is the most efficient under correctly specified nuisance functions, consistent with Theorem \ref{semi order}. However, its standard error is slightly enlarged by the outlier in the MC simulation, as indicated in Figure \ref{sim1_sd}. After removing the outlier, the standard error of MR2 under Case 1.1 (\romannumeral1) decreases to 0.14. With misspecified nuisance functions, MR2 generally has higher variance than MR1, even when Condition $2^{(1:K)}$ holds. In contrast, MR1 remains the most efficient in most cases, including those with misspecification. When only Condition $1^{(1:K)}$ holds (Case 1.2), MR2 becomes inconsistent, whereas MR1 still remains more efficient than DR-t in most cases.

\begin{figure}[htbp]
	\centering
	\begin{subfigure}{0.49\linewidth}
		\centering
		\includegraphics[width=0.95\linewidth]{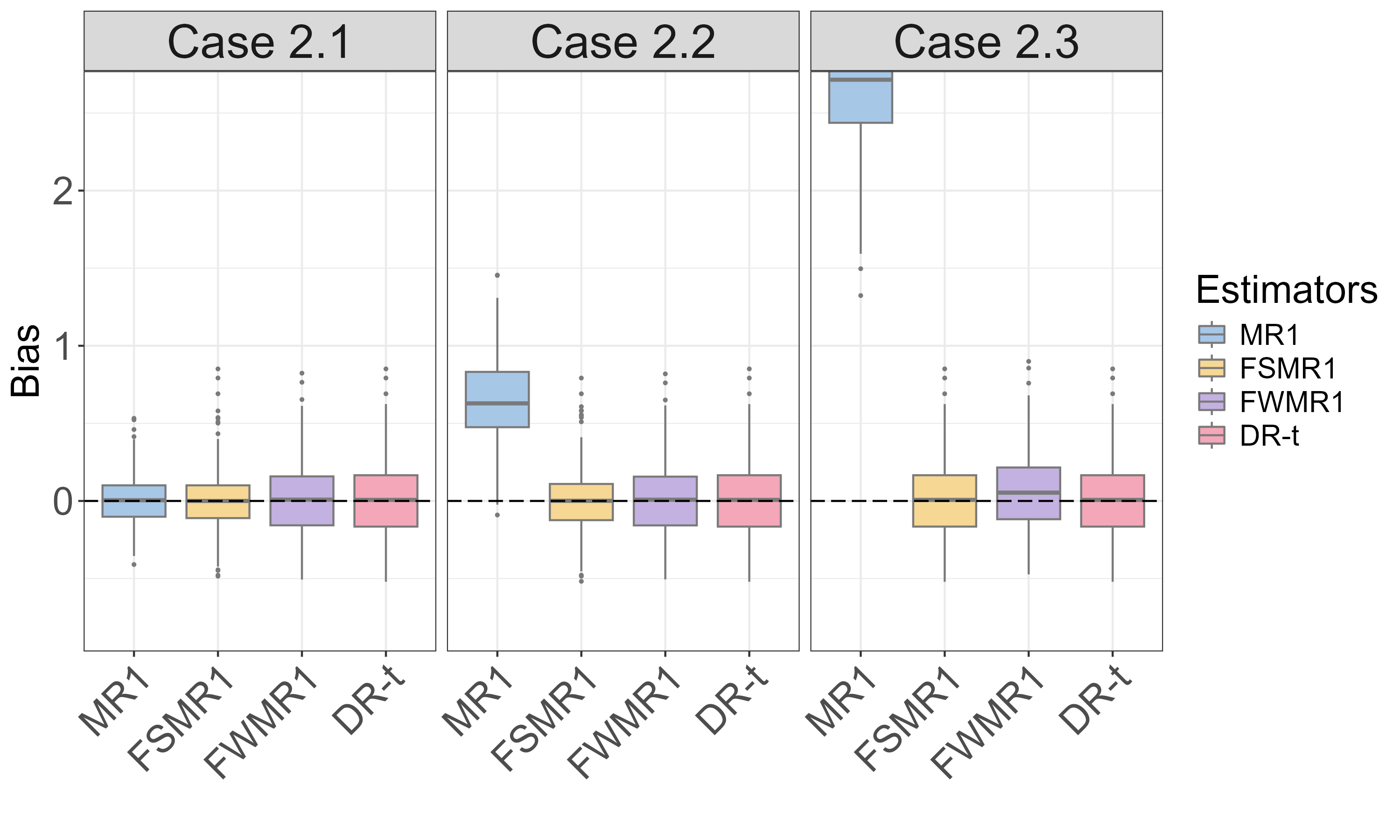}
		\caption{Bias}
		\label{sim2}
	\end{subfigure}
	\begin{subfigure}{0.49\linewidth}
		\centering
		\includegraphics[width=0.95\linewidth]{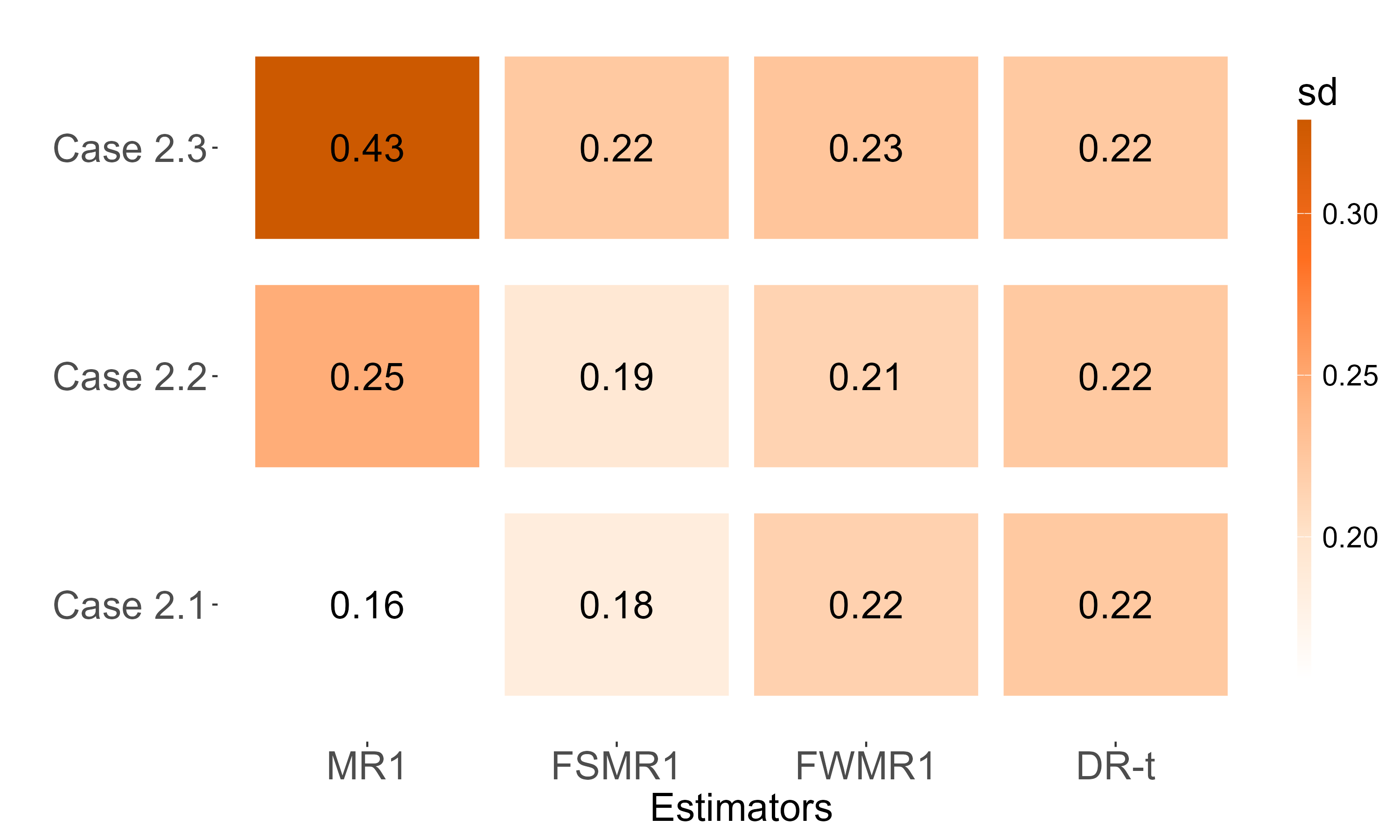}
		\caption{Standard error}
		\label{sim2_sd}
	\end{subfigure}

	\vspace{8mm} 
    
	\begin{subfigure}{0.49\linewidth}
		\centering
		\includegraphics[width=\linewidth]{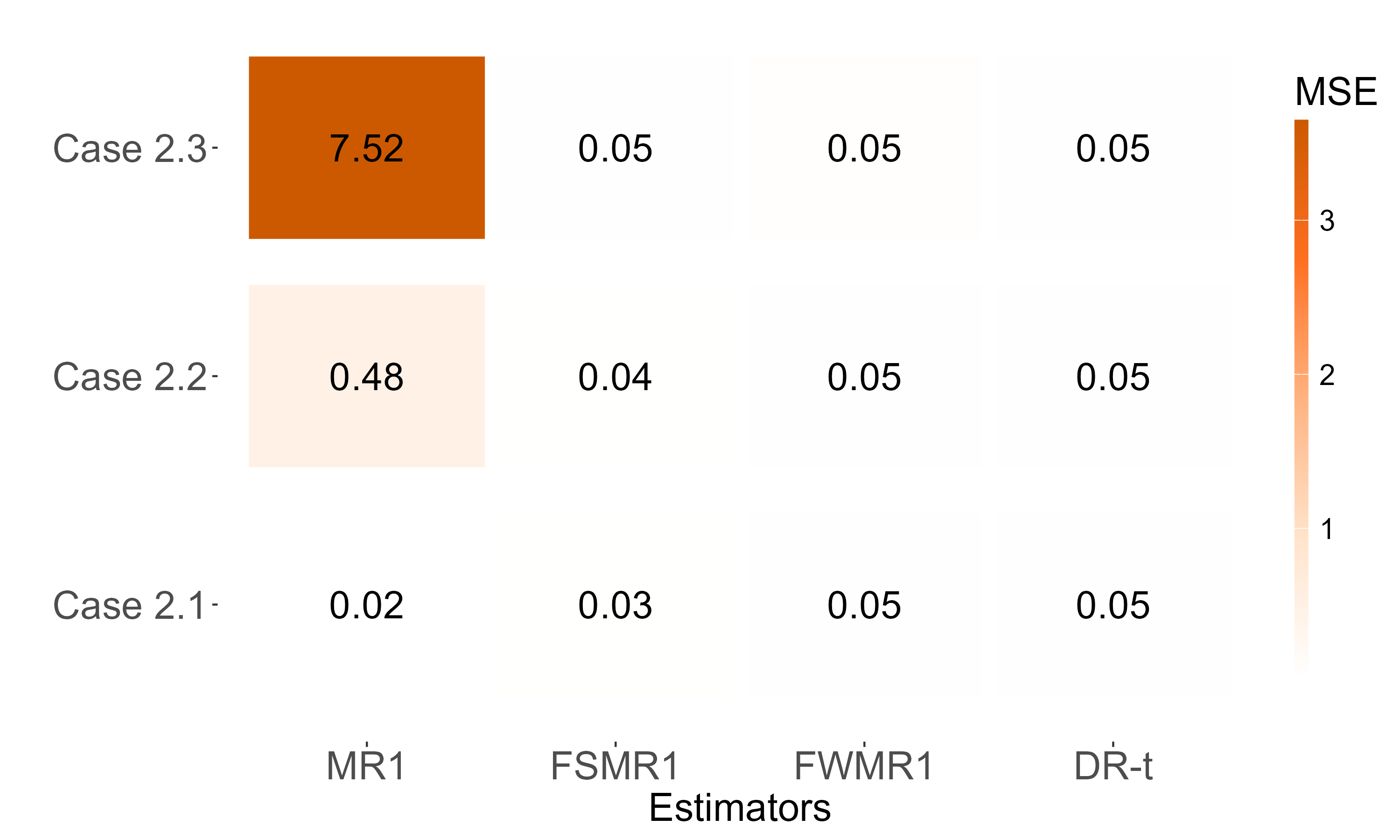}
		\caption{Mean squared error}
		\label{sim2_mse}
	\end{subfigure}
	\begin{subfigure}{0.49\linewidth}
		\centering
		\includegraphics[width=0.9\linewidth]{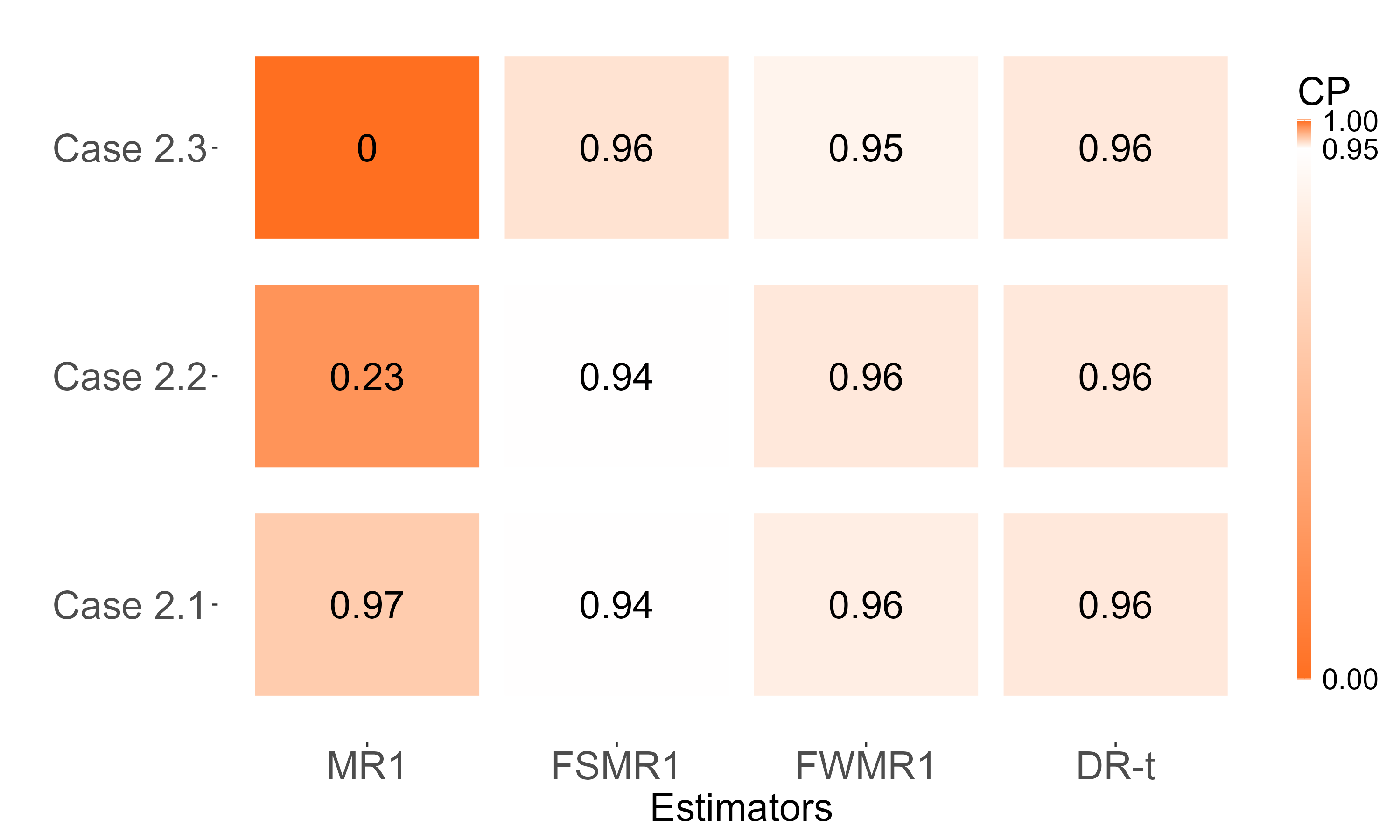}
		\caption{Coverage probability}
		\label{sim2_cp}
	\end{subfigure}

    \caption{Simulation results of Scenario 2.}
\end{figure}

In Scenario 2, we focus on varying the values of $b^{(2)}_\tau$ and $b^{(3)}_\tau$. Given the rate robustness property, flexible machine learning method can be used to estimate the nuisance functions, ensuring rate consistency in the estimation. Consequently, we will not account for misspecification in this scenario. We evaluate three levels of RR function shift (all sites transportable, some sites transportable, no sites transportable). The results are presented in Figures \ref{sim2}.

When all sites are transportable (Case 2.1), MR1 achieves the semiparametric efficiency, making it the most efficient one among these estimators. As shown in Fig. \ref{sim2}, our federated estimator FSMR1 performs similarly to MR1, demonstrating clear efficiency improvements over DR-t. Without the selection step, FWMR1 does not show substantial improvement over DR-t. When only some sites are transportable (Case 2.2), directly applying MR1 across all sites results in inconsistency. FSMR1 selectively applies MR1 to transportable sites, achieving valid coverage with shorter CI. When no sites are transportable (Case 2.3), borrowing external data is harmful. DR-t becomes the optimal choice in this case. Applying MR1 significantly increases bias. However, FSMR1 remains nearly identical to DR-t. This result establishes a lower bound for our estimator FSMR1. That is, even when no external site is transportable, FSMR1 performs no worse than using the target site data only.

\section{Real data analysis}
Accurately estimating the effect of vasopressor administration on in-hospital mortality is critical in intensive care. We apply the proposed method to a real dataset from the eICU Collaborative Research Database (eICU-CRD) \citep{goldberger2000physiobank, pollard2018eicu, pollard2019icu}, a multi-center ICU database containing de-identified health-related information from over 200,000 ICU admissions across the United States during 2014-2015.

In our analysis, the treatment variable is coded as 1 for vasopressor administration and 0 for alternative medical supervisions such as IV fluid resuscitation. The primary outcome is in-hospital mortality, coded as 1 for patients who died during hospitalization and 0 for those who survived. To account for potential confounding, we consider $p=8$ baseline covariates: age (years), admission weight (kg), temperature (Celsius), glucose level (mg/dL), blood urea nitrogen (BUN) (mg/dL), creatinine (mg/dL), white blood cell (WBC) count (K/$\mu$L), platelets (K/$\mu$L). These covariates are clinically relevant factors that influence both treatment assignment and parient outcomes. Their inclusion supports the plausibility of Assumptions \ref{unconf} and \ref{posi}. Among the available sites in the eICU-CRD, we designate site 167 as the target site, and sites 199, 243, and 449 as the source sites for potential borrowing. These four sites include a total of $n=404$ patients, with 158 patients in site 167, 76 in site 199, 84 in site 243 and 86 in site 449. Table \ref{baseline} summarizes the mean and standard deviation of each covariate and outcome across sites, revealing heterogeneity between different sites. Note that although the full dataset is available, we analyze each site separately, without pooling, to maintain the federated learning setting.

Our primary objective is to construct a CI for the RR of mortality at the target site (site 167). In our implementation, we estimated the nuisance functions using Super Learner ensemble, incorporating both generalized linear models and mean-based learners. Table \ref{real world result} reports the estimated results. Our method selects site 243 for borrowing and produces narrower CI compared to the target-only analysis, demonstrating a meaningful gain in precision. 

\begin{figure}[htbp]
    \centering
    \includegraphics[width=\linewidth]{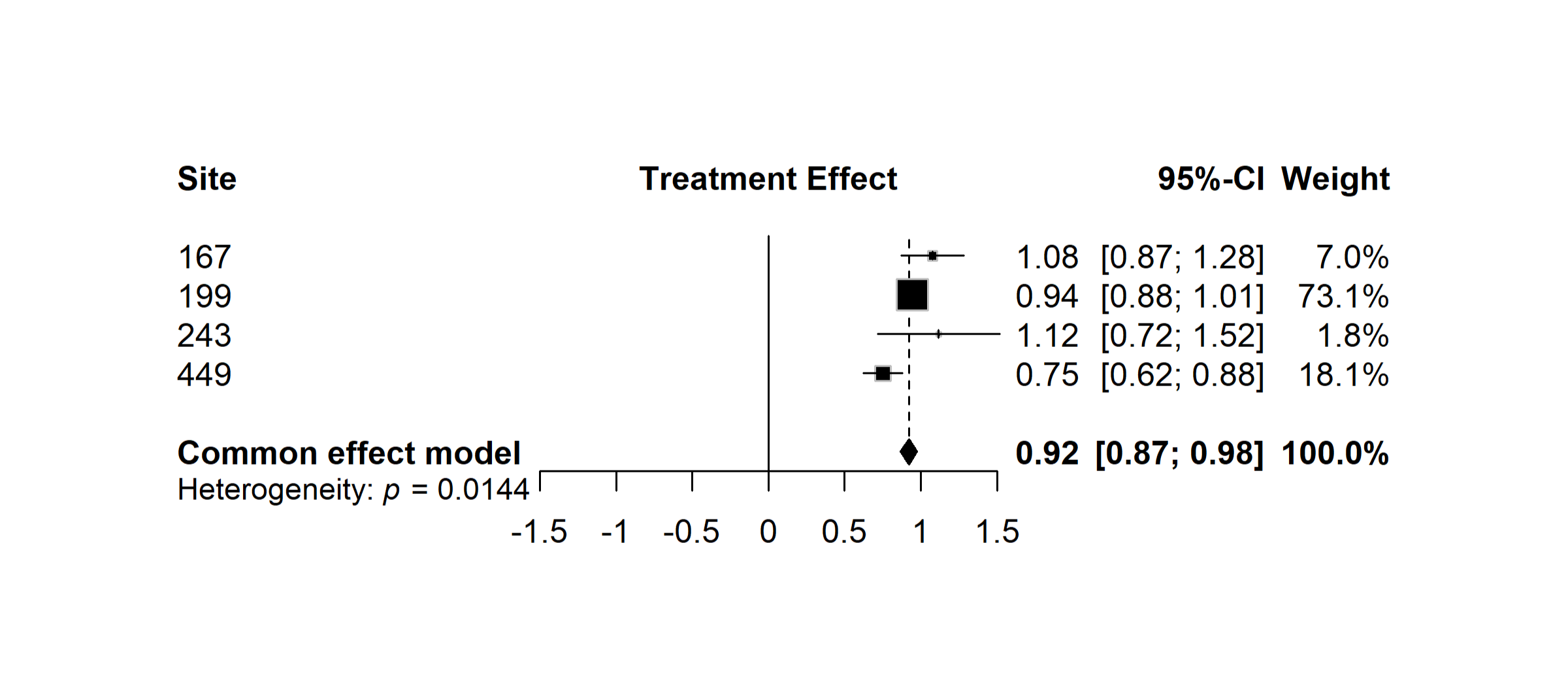}
    \caption{Black squares show site-specific treatment effects with size reflecting precision, horizontal lines indicate confidence intervals, and the black diamond shows the combined fixed-effect estimate.}
    \label{forest}
\end{figure}

To assess the validity of borrowing from site 243, we first examine the forest plot (Figure \ref{forest}), which shows that the treatment effect estimate for site 243 is close to that of the target site 167. As a supplementary diagnostic, we also compare the ratios of the $\tau(x)$ functions between the target and each source site (see Supplementary Material). Only the ratio between sites 167 and 243 is close to 1 for $\tau(x)$, further indicating compatibility with Condition \ref{con1}. Taken together, these results support the appropriateness of borrowing information from site 243 under our framework.
\begin{table}[htbp]
    \centering
    \caption{Means and standard deviations (in parenthesis) of outcome and baseline characteristics in different sites.}
    \resizebox{\textwidth}{!}{%
    \begin{tabular}{ccccc}
    & Site 167 & Site 199 & Site 243 & Site 449\\
     Mortality $(Y)$ & 0.55 (0.50) & 0.57 (0.50) & 0.49 (0.50) & 0.38 (0.49)\\
     Age $(X_1)$ & 58.16 (14.77) & 60.52 (12.37) & 62.00 (13.18) & 57.85 (13.25)\\
     Admission Weights $(X_2)$ & 87.90 (24.02) & 97.85 (25.25) & 89.23 (24.83) & 96.12 (29.64) \\
     Temperature $(X_3)$ & 36.74 (1.35) & 36.34 (1.21) & 36.24 (1.36) & 36.86 (1.28)\\
     Glucose $(X_4)$ & 150.22 (72.34) & 158.12 (77.97) & 158.60 (93.68) & 141.53 (60.21)\\
     BUN $(X_5)$ & 45.71 (29.55) & 47.32 (26.95) & 49.98 (34.64) & 47.90 (32.97)\\
     Creatinine $(X_6)$ & 2.81 (2.02) & 3.17 (1.89) & 3.60 (2.79) & 3.27 (2.43) \\
     WBC $(X_7)$ & 14.61 (9.47) & 17.22 (11.11) & 16.54 (10.09) & 14.92 (11.70)\\
     Platelets $(X_8)$ & 154.28 (103.65) & 162.75 (94.11) & 182.67 (125.58) & 164.94 (142.36)
    \end{tabular}
    }
    \label{baseline}
\end{table}
\begin{table}[htbp]
    \centering
    \caption{SE, standard error; CI, confidence interval.}
    \begin{tabular}{cccc}
     & $\widehat\tau^{(0)}$ & $\widehat{SE}(\widehat\tau^{(0)})$ & $95\%$ CI\\
     DR-t & 1.08 & 0.32 & $[0.44, 1.71]$\\
     FSMR1 & 0.97 & 0.22 & $[0.54, 1.39]$\\
    \end{tabular} 
    \label{real world result}
\end{table}


\section{Discussion}
Rather than relying on selection to mitigate biases when borrowing external information, \citet{yang2025data} demonstrates that explicitly modeling and correcting the bias function can be an effective strategy, and extending this approach to federated settings is a promising direction for future work. Additionally, federated learning involving small sample sizes \citep{zhu2024enhancing} demands fundamentally different considerations, where methods such as randomization-based inference may be more appropriate.

\section{Acknowledgments}
We thank Zehao Sun and Frank Eriksson for their careful review of an earlier version of this manuscript.

\bibliographystyle{apalike}  
\bibliography{refs}  

@article{colnet2023risk,
  title={Risk ratio, odds ratio, risk difference... Which causal measure is easier to generalize?},
  author={Colnet, B{\'e}n{\'e}dicte and Josse, Julie and Varoquaux, Ga{\"e}l and Scornet, Erwan},
  journal={arXiv preprint arXiv:2303.16008},
  year={2023}
}

@article{liu2024multi,
  title={Multi-source conformal inference under distribution shift},
  author={Liu, Yi and Levis, Alexander W and Normand, Sharon-Lise and Han, Larry},
  journal={Proceedings of machine learning research},
  volume={235},
  pages={31344},
  year={2024}
}

@article{gao2025improving,
  title={Improving randomized controlled trial analysis via data-adaptive borrowing},
  author={Gao, Chenyin and Yang, Shu and Shan, Mingyang and Ye, Wenyu and Lipkovich, Ilya and Faries, Douglas},
  journal={Biometrika},
  volume={112},
  number={2},
  pages={asae069},
  year={2025},
  publisher={Oxford University Press}
}

@article{xiong2023federated,
  title={Federated causal inference in heterogeneous observational data},
  author={Xiong, Ruoxuan and Koenecke, Allison and Powell, Michael and Shen, Zhu and Vogelstein, Joshua T and Athey, Susan},
  journal={Statistics in Medicine},
  volume={42},
  number={24},
  pages={4418--4439},
  year={2023},
  publisher={Wiley Online Library}
}

@book{hartung2011statistical,
  title={Statistical meta-analysis with applications},
  author={Hartung, Joachim and Knapp, Guido and Sinha, Bimal K},
  year={2011},
  publisher={John Wiley \& Sons}
}

@article{han2023multiply,
  title={Multiply robust federated estimation of targeted average treatment effects},
  author={Han, Larry and Shen, Zhu and Zubizarreta, Jose},
  journal={Advances in Neural Information Processing Systems},
  volume={36},
  pages={70453--70482},
  year={2023}
}

@article{moher2010consort,
  title={CONSORT 2010 explanation and elaboration: updated guidelines for reporting parallel group randomised trials},
  author={Moher, David and Hopewell, Sally and Schulz, Kenneth F and Montori, Victor and G{\o}tzsche, Peter C and Devereaux, Philip J and Elbourne, Diana and Egger, Matthias and Altman, Douglas G},
  journal={Bmj},
  volume={340},
  year={2010},
  publisher={British Medical Journal Publishing Group}
}

@article{naylor1992measured,
  title={Measured enthusiasm: does the method of reporting trial results alter perceptions of therapeutic effectiveness?},
  author={Naylor, C David and Chen, Erluo and Strauss, Barry},
  journal={Annals of Internal Medicine},
  volume={117},
  number={11},
  pages={916--921},
  year={1992},
  publisher={American College of Physicians}
}

@article{forrow1992absolutely,
  title={Absolutely relative: how research results are summarized can affect treatment decisions},
  author={Forrow, Lachlan and Taylor, William C and Arnold, Robert M},
  journal={The American journal of medicine},
  volume={92},
  number={2},
  pages={121--124},
  year={1992},
  publisher={Elsevier}
}

@book{rothman2024epidemiology,
  title={Epidemiology: an introduction},
  author={Rothman, Kenneth J and Huybrechts, Krista F and Murray, Eleanor J},
  year={2024},
  publisher={Oxford university press}
}

@article{pollard2019icu,
  title={ICU Collaborative Research Database (Version 2.0)},
  author={Pollard, Tom and Johnson, Alistair and Raffa, Jesse and Celi, Leo Anthony and Badawi, Omar and Mark, Roger},
  journal={PhysioNet. Published online},
  volume={10},
  pages={C2WM1R},
  year={2019}
}

@article{pollard2018eicu,
  title={The eICU Collaborative Research Database, a freely available multi-center database for critical care research},
  author={Pollard, Tom J and Johnson, Alistair EW and Raffa, Jesse D and Celi, Leo A and Mark, Roger G and Badawi, Omar},
  journal={Scientific data},
  volume={5},
  number={1},
  pages={1--13},
  year={2018},
  publisher={Nature Publishing Group}
}

@article{goldberger2000physiobank,
  title={PhysioBank, PhysioToolkit, and PhysioNet: components of a new research resource for complex physiologic signals},
  author={Goldberger, Ary L and Amaral, Luis AN and Glass, Leon and Hausdorff, Jeffrey M and Ivanov, Plamen Ch and Mark, Roger G and Mietus, Joseph E and Moody, George B and Peng, Chung-Kang and Stanley, H Eugene},
  journal={circulation},
  volume={101},
  number={23},
  pages={e215--e220},
  year={2000},
  publisher={Lippincott Williams \& Wilkins}
}

@article{yang2023elastic,
  title={Elastic integrative analysis of randomised trial and real-world data for treatment heterogeneity estimation},
  author={Yang, Shu and Gao, Chenyin and Zeng, Donglin and Wang, Xiaofei},
  journal={Journal of the Royal Statistical Society Series B: Statistical Methodology},
  volume={85},
  number={3},
  pages={575--596},
  year={2023},
  publisher={Oxford University Press US}
}

@article{su2024efficient,
  title={Efficient estimation of the target population average treatment effect from multi-source data},
  author={Su, Zehao and Rytgaard, Helene Charlotte and Ravn, Henrik and Eriksson, Frank},
  journal={arXiv preprint arXiv:2405.10769},
  year={2024}
}

@article{li2024efficient,
  title={Efficient combination of observational and experimental datasets under general restrictions on outcome mean functions},
  author={Li, Harrison H},
  journal={arXiv e-prints},
  pages={arXiv--2406},
  year={2024}
}

@article{yang2025data,
  title={Data fusion methods for the heterogeneity of treatment effect and confounding function},
  author={Yang, Shu and Liu, Siyi and Zeng, Donglin and Wang, Xiaofei},
  journal={Bernoulli},
  volume={31},
  number={4},
  pages={2987--3012},
  year={2025},
  publisher={Bernoulli Society for Mathematical Statistics and Probability}
}

@article{gao2023pretest,
  title={Pretest estimation in combining probability and non-probability samples},
  author={Gao, Chenyin and Yang, Shu},
  journal={Electronic Journal of Statistics},
  volume={17},
  number={1},
  pages={1492--1546},
  year={2023},
  publisher={The Institute of Mathematical Statistics and the Bernoulli Society}
}

@inproceedings{wu2022integrative,
  title={Integrative $ R $-learner of heterogeneous treatment effects combining experimental and observational studies},
  author={Wu, Lili and Yang, Shu},
  booktitle={Conference on Causal Learning and Reasoning},
  pages={904--926},
  year={2022},
  organization={PMLR}
}

@article{cheng2024inference,
  title={Inference for optimal linear treatment regimes in personalized decision-making},
  author={Cheng, Yuwen and Yang, Shu},
  journal={arXiv preprint arXiv:2405.16161},
  year={2024}
}

@article{zhu2024enhancing,
  title={Enhancing statistical validity and power in hybrid controlled trials: A randomization inference approach with conformal selective borrowing},
  author={Zhu, Ke and Yang, Shu and Wang, Xiaofei},
  journal={arXiv preprint arXiv:2410.11713},
  year={2024}
}

@book{bickel1993efficient,
  title={Efficient and adaptive estimation for semiparametric models},
  author={Bickel, Peter J and Klaassen, Chris AJ and Bickel, Peter J and Ritov, Ya’acov and Klaassen, J and Wellner, Jon A and Ritov, YA'Acov},
  volume={4},
  year={1993},
  publisher={Springer}
}

@article{jiang2022multiply,
  title={Multiply robust estimation of causal effects under principal ignorability},
  author={Jiang, Zhichao and Yang, Shu and Ding, Peng},
  journal={Journal of the Royal Statistical Society Series B: Statistical Methodology},
  volume={84},
  number={4},
  pages={1423--1445},
  year={2022},
  publisher={Oxford University Press}
}

@book{van2000asymptotic,
  title={Asymptotic statistics},
  author={Van der Vaart, Aad W},
  volume={3},
  year={2000},
  publisher={Cambridge university press}
}

@misc{chernozhukov2018double,
  title={Double/debiased machine learning for treatment and structural parameters},
  author={Chernozhukov, Victor and Chetverikov, Denis and Demirer, Mert and Duflo, Esther and Hansen, Christian and Newey, Whitney and Robins, James},
  year={2018},
  publisher={Oxford University Press Oxford, UK}
}

@article{han2025federated,
  title={Federated adaptive causal estimation (face) of target treatment effects},
  author={Han, Larry and Hou, Jue and Cho, Kelly and Duan, Rui and Cai, Tianxi},
  journal={Journal of the American Statistical Association},
  pages={1--14},
  year={2025},
  publisher={Taylor \& Francis}
}
\newpage
\section*{Supplementary Material}
All technical details are provided in this Supplementary Material. Specifically, $\S$\ref{proof} presents the general procedure for obtaining the EIF for various causal measures, along with the proofs of Theorems 1–7, $\S$\ref{dgp} gives the data generating process of our simulation studies, and $\S$\ref{realdata} presents additional diagnostics assessing the reasonableness of the sites selected by our method when applied to real data.

\section{Technical proofs and details}
\label{proof}
\subsection{Efficient influence functions for various causal measures: a general procedure}
Under Condition $1^{(1:K)}$, the causal measure satisfying Assumption 3 in the target site (i.e., site 0) can be identified using the combined data across all sites
$$\psi^{(0)}=m(\psi_0^{(0)},\psi_1^{(0)})
    =m(\mathbb{E}\{\mathbb{E}(Y|X,A=0,S=0)|S=0\},\mathbb{E}\{\mathbb{E}(Y|X,A=1,S=0)|S=0\}).$$

As noted in the main text, it suffices to derive the Efficient Influence Function (EIF) for $\psi_1^{(0)}=\mathbb{E}\{\mathbb{E}(Y|X,A=1,S=0)|S=0\}$. Equivalently, $\psi_1^{(0)}$ can be expressed in the following form:
\begin{align}
    \psi_{1}^{(0)}&=\mathbb{E}\{\mathbb{E}(Y|X,A=1,S=0)|S=0\}\nonumber\\
    &=\mathbb{E}\{I(S=0)\mathbb{E}(Y|X,A=1,S=0)\}/\operatorname{pr}(S=0)\nonumber\\
    &=\mathbb{E}\{\operatorname{pr}(S=0|X)\mu_{1}^{(0)}(X)\}/\operatorname{pr}(S=0).
    \label{psi1}
\end{align}
Let $V=(Y,X,A,S)$ be a vector of random variables, the joint pdf from one single observation is
\begin{align*}
    f(Y,X,A,S)&=f(X)\\
            &\times	\prod_{k=0}^{K}\operatorname{pr}(S=k|X)^{I(S=k)}\\
            &\times	\prod_{k=0}^{K}\left\{\operatorname{pr}(A=1|S=k,X)^{A}\operatorname{pr}(A=0|S=k,X)^{1-A}\right\}^{I(S=k)}\\
            &\times	\prod_{k=0}^{K}\left\{f(Y|X,S=k,A=1)^{A}f(Y|X,S=k,A=0)^{1-A}\right\}^{I(S=k)}.
\end{align*}
To derive the EIF, we employ the parametric submodel approach of \citet{bickel1993efficient}. Let $\{f_t(V):t\in\mathbb R\}$ denote a regular parametric submodel, with the true distribution corresponding to $t=0$, i.e., $f_t(V)|_{t=0}=f(V)$. Then, the tangent space $\mathcal{T}$ can be constructed by $\mathcal{T}=\mathcal{T}_{1}+\mathcal{T}_{2}+\mathcal{T}_{3}+\mathcal{T}_{4}$, where

$\mathcal{T}_{1}=\left\{a(X):\int a(X)f(X)dx=0\right\}$

$\mathcal{T}_{2}=\{b(X,S):\sum_{k=0}^Kb(X,S=k)\operatorname{pr}(S=k|X)=0\}$

$\mathcal{T}_{3}=\{c(X,S,A):\sum_{a=0,1}c(X,S,a)\operatorname{pr}(A=a|X,S)=0\}$

\begin{align*}
    \mathcal{T}_{4}	=\mathcal{T}_{41}\cap\mathcal{T}_{42}&=\left\{d(Y,X,S,A):\int d(y,X,S,A)f(y|X,S,A)dy=0\right\} \\
\cap_{k=1}^K	&\left\{d(Y,X,S,A): \mathbb{E}\{\eta_1^{(k)}(V)d(Y,X,S,A)|X\}=0\right\},
\end{align*}
where $\eta_1^{(k)}(V)$ satisfies the restriction introduced below for all score functions $S(Y,A,S|X)$.

To obtain an expression of $\eta_1^{(k)}$, we first recall the following notations:
$$\mu_{1}^{(k)}(x)=g_{\mu_{0}^{(k)}(x)}^{-1}(\tau^{(k)}(x))\quad\text{and}\quad \tau^{(k)}(x)=g_{\mu_{0}^{(k)}(x)}(\mu_{1}^{(k)}(x)),\quad k\in[[K]].$$

Then the restriction $\tau^{(0)}(X)=\cdots=\tau^{(K)}(X)$ can be equivalently written as:
\begin{equation}
\label{genres}
    g_{\mu_{0}^{(0)}(x)}(\mu_{1}^{(0)}(x))-g_{\mu_{0}^{(k)}(x)}(\mu_{1}^{(k)}(x))=0,\quad k\in[[K]]\backslash\{0\}.
\end{equation}

Taking the pathwise derivative on both sides of Equation \ref{genres} and applying Equations \ref{mu1} and \ref{mu0} (introduced below), there exists a function $\eta_1^{(k)}(V)$ such that the resulting expression can be written as $\mathbb{E}\{\eta_1^{(k)}(V)S(Y,A,S|X)|X\}=0$.

Denote the EIF for $\psi^{(0)}_{1}$ as $\varphi^{(0)}_{1}$, which must satisfy
$(\partial \psi_{1,t}^{(0)}/\partial t)|_{t=0}=\mathbb{E}[\varphi_{1}^{(0)}S(V)]$ and belongs to the tangent space $\mathcal{T}$. Here $S(V)$ denotes the observed score function. From Equation \ref{psi1}, $\psi_1^{(0)}$ can be expressed as a ratio with numerator $N_{1}^{(0)}=\mathbb{E}\{\operatorname{pr}(S=0|X)\mu_{1}^{(0)}(X)\}$ and denominator $D_{1}^{(0)}=\operatorname{pr}(S=0)$. Let $N_{1,t}^{(0)}$ and $D_{1,t}^{(0)}$ denote the corresponding quantities evaluated under the parametric submodel $f_t(V)$.

For the numerator $N_{1}^{(0)}$, the semiparametric efficiency score is the pathwise derivative, which is:
\begin{align*}
    \left.\frac{\partial N_{1,t}^{(0)}}{\partial t}\right|_{t=0}	&=\mathbb{\mathbb{E}}\{\operatorname{pr}(S=0|X)\mu_{1}^{(0)}(X)S(X)\}\\
    &+\mathbb{E}_{t}\left.\left\{\frac{\partial \operatorname{pr}(S=0|X)}{\partial t}\mu_{1,t}^{(0)}(X)\right\}\right|_{t=0}+\mathbb{E}_{t}\left.\left\{\operatorname{pr}(S=0|X)\frac{\partial\mu_{1,t}^{(0)}(X)}{\partial t}\right\}\right|_{t=0}.
\end{align*}

We now calculate the pathwise derivative in the second part:
$$
    \left.\frac{\partial \operatorname{pr}(S=0|X)}{\partial t}\right|_{t=0}=\left.\frac{\partial}{\partial t}\mathbb{E}_{t}\{I(S=0)|X\}\right|_{t=0}
    =\mathbb{E}[\{I(S=0)-\operatorname{pr}(S=0|X)\}S(Y,A,S|X)|X].
$$

Then we calculate the pathwise derivative in the third part. There exist $K+1$ distinct ways to calculate $\partial\mu_{1,t}^{(0)}(X)/\partial t$ under Condition $1^{(1:K)}$, i.e. $\tau^{(0)}(X)=\cdots=\tau^{(K)}(X)$.

First, $\partial\mu_{1,t}^{(0)}(X)/\partial t$ can be computed using data from site 0 only:
\begin{align*}
    \left.\frac{\partial\mu_{1,t}^{(0)}(X)}{\partial t}\right|_{t=0}&=\left.\left\{\int y\frac{\partial}{\partial t}f_{t}(y|X,A=1,S=0)dy\right\}\right|_{t=0}\\
    &=\mathbb{E}\left\{YS(Y|X,A=1,S=0)|X,A=1,S=0\right\}\\
    &=\mathbb{E}\left[\{Y-\mu_{1}^{(0)}(X)\}S(Y|X,A=1,S=0)|X,A=1,S=0\right]\\
    &=\mathbb{E}\left[\frac{AI(S=0)\{Y-\mu_{1}^{(0)}(X)\}S(Y,A,S|X)}{\operatorname{pr}(S=0|X)\pi^{(0)}(X)}\bigg| X\right]\\
    &=\mathbb{E}\left[H_1^{(0)}(V)S(Y,A,S|X)/\operatorname{pr}(S=0|X)| X\right],
\end{align*}
where $H_1^{(0)}(V)=AI(S=0)\{Y-\mu_{1}^{(0)}(X)\}/\pi^{(0)}(X)$.

Similarly, we can obtain the pathwise derivative of $\partial\mu_{a,t}^{(k)}(X)/\partial t$ for $a=0,1$ and each site $k\in[[K]]$:
\begin{equation}
\label{mu1}
\left.\frac{\partial\mu_{1,t}^{(k)}(X)}{\partial t}\right|_{t=0}=\mathbb{E}\left[\left.\frac{AI(S=k)\{Y-\mu_{1}^{(k)}(X)\}S(Y,A,S|X)}{\operatorname{pr}(S=k|X)\pi^{(k)}(X)}\right|X\right],
\end{equation}
and
\begin{equation}
\label{mu0}
    \left.\frac{\partial\mu_{0,t}^{(k)}(X)}{\partial t}\right|_{t=0}=\mathbb{E}\left[\left.\frac{(1-A)I(S=k)\{Y-\mu_{0}^{(k)}(X)\}S(Y,A,S|X)}{\operatorname{pr}(S=k|X)\{1-\pi^{(k)}(X)\}}\right|X\right].
\end{equation}

The remaining $K$ approaches for computing $\partial\mu_{1,t}^{(0)}(X)/\partial t$ each rely on one of the $K$ restrictions in Condition $1^{(1:K)}$. For instance, using the assumption $\tau^{(k)}(X)=\tau^{(0)}(X)$ for $k\neq 0$, we can express $\mu_{1}^{(0)}(X)$ as follows:
\begin{align*}
    \mu_{1}^{(0)}(X) &= g_{\mu_{0}^{(0)}(X)}^{-1}(\tau^{(0)}(X))\\
    &= g_{\mu_{0}^{(0)}(X)}^{-1}(\tau^{(k)}(X))\\
    &= g_{\mu_{0}^{(0)}(X)}^{-1}(g_{\mu_{0}^{(k)}(X)}(\mu_{1}^{(k)}(X))).
\end{align*}

Let $G^{(k)}:=G^{(k)}(\mu_{0}^{(0)}(X), \mu_{0}^{(k)}(X),\mu_1^{(k)}(X))=g_{\mu_{0}^{(0)}(X)}^{-1}\circ g_{\mu_{0}^{(k)}(X)}\circ\mu_{1}^{(k)}(X)$. Under the restriction $\tau^{(k)}(X)=\tau^{(0)}(X)$, the pathwise derivative of $\mu_{1}^{(0)}(X)$ can then be expressed as:
$$\left.\frac{\partial\mu_{1,t}^{(0)}(X)}{\partial t}\right|_{t=0}	= \left.\frac{\partial G^{(k)}_t}{\partial t}\right|_{t=0},\quad k\in[[K]]\backslash\{0\}.$$

Notice that when taking pathwise derivative, we need to guarantee that $\mathbb E\{H_1^{(k)}\}=0$. Based on Equation \ref{mu1} and Equation \ref{mu0}, we can write the above pathwise derivative as the form:
$$\mathbb{E}\{H_1^{(k)}(V)S(Y,A,S|X)/\operatorname{pr}(S=0|X)|X\}.$$

We combine these $K+1$ derivatives as follows:
$$\left.\frac{\partial\mu_{1,t}^{(0)}(X)}{\partial t}\right|_{t=0}=\mathbb{E}\left\{\sum_{k=0}^K\frac{C_{1k}}{\sum_{j=0}^KC_{1j}}H_1^{(k)}(V)S(Y,A,S|X)/\operatorname{pr}(S=0|X)\bigg|X\right\},$$
where $C_k=C_k(X),k\in[[K]]$ are $K+1$ arbitrary function of $X$. To obtain the EIF of $N_1^{(0)}$, we need to find the proper $C_k$ such that the third term belongs to the tangent space. That is,

\begin{align}
\label{restriction}
    \mathbb{E}\left\{\eta_1^{(l)}(V)\sum_{k=0}^KC_{1k}H_1^{(k)}(V)\bigg|X\right\}=0,\quad l\in[[K]]\backslash\{0\}.
\end{align}

By solving the system of $K$ equations, we obtain the exact forms of $C_{10}/C_{1k}$. It is sufficient to calculate the EIF for $N_{1}^{(0)}$, which is
\begin{align*}
    \varphi_{1N}^{(0)}&=\operatorname{pr}(S=0|X)\mu_{1}^{(0)}(X)-N_{1}^{(0)}\\
    &+\{I(S=0)-\operatorname{pr}(S=0|X)\}\mu_{1}^{(0)}(X)
    +\sum_{k=0}^K\frac{C_{1k}}{\sum_{j=1}^{K}C_{1j}}H_1^{(k)}(V).
\end{align*}

Back to the denominator of $\psi_{1}^{(0)}$, we can show that $\varphi_{1D}^{(0)}=I(S=0)-D_{1}^{(0)}$ is its efficient score. Then, the semiparametric efficient score for $\psi_1^{(0)}$ based on Lemma S2 in \citet{jiang2022multiply} is
$$\varphi_{1}^{(0)}(V)=\frac{I(S=0)\left\{\mu_1^{(0)}(X)-\psi^{(0)}_{1}\right\}+\sum_{k=0}^K\frac{C_{1k}}{\sum_{j=0}^{K}C_{1j}}H_1^{(k)}(V)}{\operatorname{pr}(S=0)}.$$

From above, by setting the empirical efficient score under the observed data $V=(Y,X,A,S)$ equals to zero, the semiparametric efficient estimator of $\psi_{1}^{(0)}$ is obtained by
$$\widehat{\psi}_{1}^{(0)}=\frac{\sum_{i=1}^n\left\{I(S_i=0)\widehat\mu_1^{(0)}(X_i)+\sum_{k=0}^K\frac{C_{1k}(X_i)}{\sum_{j=1}^{K}C_{1j}(X_i)}\widehat H_1^{(k)}(V_i)\right\}}{\sum_{i=1}^nI(S_i=0)}.
$$

\subsection{Proof of Theorem 1}
To identify the Risk Ratio (RR) in the target site, i.e. the site $0$, with the data set across all the sites, we have
$$\psi^{(0)}=\frac{\mathbb{E}(Y^{(1)}|S=0)}{\mathbb{E}(Y^{(0)}|S=0)}=\frac{\mathbb{E}\{\mathbb{E}(Y|X,A=1,S=0)|S=0\}}{\mathbb{E}\{\mathbb{E}(Y|X,A=0,S=0)|S=0\}}.$$

In this case, we can specifically calculate that
\begin{align*}
    \eta_1^{(k)}(V)&=\frac{I(S=0)A}{\operatorname{pr}(S=0|X)\pi^{(0)}(X)}\frac{Y}{\mu_{0}^{(0)}(X)}-\frac{I(S=0)(1-A)}{\operatorname{pr}(S=0|X)\{1-\pi^{(0)}(X)\}}\frac{Y\mu_{1}^{(0)}(X)}{\{\mu_{0}^{(0)}(X)\}^{2}}\\
&-\frac{I(S=k)A}{\operatorname{pr}(S=k|X)\pi^{(k)}(X)}\frac{Y}{\mu_{0}^{(k)}(X)}+\frac{I(S=k)(1-A)}{\operatorname{pr}(S=k|X)\{1-\pi^{(k)}(X)\}}\frac{Y\mu_{1}^{(k)}(X)}{\{\mu_{0}^{(k)}(X)\}^{2}}
\end{align*}
and 
\begin{align*}
    H_1^{(k)}(V)&=\frac{I(S=k)Aq^{(k)}(X)}{\pi^{(k)}(X)}\frac{\{Y-\mu_{1}^{(k)}(X)\}\mu_{0}^{(0)}(X)}{\mu_{0}^{(k)}(X)}\\
    &-\frac{I(S=k)(1-A)q^{(k)}(X)}{1-\pi^{(k)}(X)}\frac{\{Y-\mu_{0}^{(k)}(X)\}\mu_{1}^{(0)}(X)}{\mu_{0}^{(k)}(X)}\\
    &+\frac{I(S=0)(1-A)}{1-\pi^{(0)}(X)}\frac{\mu_{1}^{(k)}(X)\{Y-\mu_{0}^{(0)}(X)\}}{\mu_{0}^{(k)}(X)},
\end{align*}
where $q^{(k)}(X)=\operatorname{pr}(S=0|X)/\operatorname{pr}(S=k|X)$.

We want to combine these $K+1$ derivatives as,
$$\left.\frac{\partial\mu_{1,t}^{(0)}(X)}{\partial t}\right|_{t=0}=\mathbb{E}\left\{\sum_{k=1}^K\frac{C_{1k}}{\sum_{k=1}^KC_{1k}}H_1^{(k)}(V)S(Y,A,S|X)/\operatorname{pr}(S=0|X)\bigg|X\right\}$$
where $C_{1k}=C_{1k}(X),k\in[[K]]$ are $K+1$ arbitrary function of $X$.

By solving the system of $K$ equations implied by restrictions \ref{restriction}, we can obtain that,
\begin{align*}
    \frac{C_{10}}{C_{1k}}&=\frac{q^{(k)}(X)\pi^{(0)}(X)\sigma^2_{1,k}(X)}{\pi^{(k)}(X)\sigma^2_{1,0}(X)}\left\{\frac{\mu_{0}^{(0)}(X)}{\mu_{0}^{(k)}(X)}\right\}^{2}
    +\frac{q^{(k)}(X)\pi^{(0)}(X)\mu_{0}^{(0)}(X)\sigma^2_{0,k}(X)}{\{1-\pi^{(k)}(X)\}\sigma^2_{1,0}(X)}\left\{\frac{\mu_{0}^{(0)}(X)\tau(X)}{\mu_{0}^{(k)}(X)}\right\}^{2}\\
    &+\frac{\pi^{(0)}(X)\sigma^2_{0,0}(X)\{\tau(X)\}^{2}}{\{1-\pi^{(0)}(X)\}\sigma^2_{1,0}(X)}\sum_{l\neq 0}\frac{g^{(k)}(X)}{g^{(l)}(X)}\quad \text{for } k\in[[K]]\backslash\{0\},
\end{align*}
where $\sigma^2_{a,k}(X)=\operatorname{var}(Y|A=a,S=k,X)$ for $k\in[[K]],\ a=0,1$, and
$g^{(k)}(X)=[\sigma^2_{1,k}(X)/\pi^{(k)}(X)+\sigma^2_{0,k}(X)\{\tau(X)\}^2/\{1-\pi^{(k)}(X)\}]/[\operatorname{pr}(S=k|X)\{\mu_{0}^{(k)}(X)\}^{2}].$

Then, the EIF for $\psi_1^{(0)}$ is
$$\varphi_{1}^{(0)}(V)=\frac{I(S=0)\left\{\mu_1^{(0)}(X)-\psi^{(0)}_{1}\right\}+\sum_{k=0}^K\frac{C_{1k}}{\sum_{j=0}^{K}C_{1j}}H_1^{(k)}(V)}{\operatorname{pr}(S=0)}.$$

\subsection{Proof of Theorem 2}
In this case, the tangent space must be redefined, and the third term of the pathwise derivative of $N_{1}^{(0)}$ should be recalculated. 

The first three components of the tangent space remain unchanged, while the fourth component, under Condition $2^{(1:K)}$, becomes
\begin{align*}
    \mathcal{T}_{4}	&=\mathcal{T}_{41}\cap\mathcal{T}_{42}=\left\{d(Y,X,S,A):\int d(y,X,S,A)f(y|X,S,A)dy=0\right\}\\
    &\cap_{k=1}^K	\left\{d(Y,X,S,A):\mathbb{E}\left\{\left[\frac{I(S=0)AY}{\operatorname{pr}(S=0|X)\pi^{(0)}(X)}-\frac{I(S=k)AY}{\operatorname{pr}(S=k|X)\pi^{(k)}(X)}\right]d(Y,X,S,A)\bigg|X\right\}=0\right\}
\end{align*}

The first approach to calculating $\partial\mu_{1,t}^{(0)}(X)/\partial t$ remains unchanged, whereas the remaining $K$ approaches are given by:
\begin{align*}
    \left.\frac{\partial\mu_{1,t}^{(0)}(X)}{\partial t}\right|_{t=0}&=\left.\frac{\partial\mu_{1,t}^{(k)}(X)}{\partial t}\right|_{t=0}\\
    &=\mathbb{E}\left.\left[\left\{Y-\mu_{1}^{(k)}(X)\right\}S(Y|X,A=1,S=k)\right|X,A=1,S=k\right]\\
    &=\mathbb{E}\left[\left.\frac{AI(S=k)\{Y-\mu_{1}^{(0)}(X)\}S(Y,A,S|X)}{\operatorname{pr}(S=k|X)\pi^{(k)}(X)}\right|X\right],\quad k\in[[K]]\backslash\{0\}.
\end{align*}
Here we denote
$$H_2^{(k)}(V)=\frac{AI(S=k)q^{(k)}(X)\{Y-\mu_{1}^{(0)}(X)\}}{\pi^{(k)}(X)}.$$

By using the same technique of combining the $K+1$ derivatives, we obtain that 
$$\frac{C_{20}}{C_{2k}}=\frac{C_{20}(X)}{C_{2k}(X)}=\frac{q^{(k)}(X)\pi^{(0)}(X)\sigma^2_{1,k}(X)}{\{\pi^{(k)}(X)\sigma^2_{1,0}(X)\}},$$
where $\sigma^2_{a,k}(X)=\operatorname{var}(Y|A=a,S=k,X)$ for $k\in[[K]],\ a=0,1$.

Then the efficient score for $N_{1}^{(0)}$ under Condition $2^{(1:K)}$ is:
\begin{align*}
    \varphi_{1N}^{(0)}&=f(S=0|X)\mu_{1}^{(0)}(X)-N_{1}^{(0)}\\
    &+\{I(S=0)-f(S=0|X)\}\mu_{1}^{(0)}(X)
    +\sum_{k=0}^K\frac{C_{2k}}{\sum_{j=1}^{K}C_{2j}}H_2^{(k)}(V).
\end{align*}

The efficient score for $\psi_1^{(0)}$ is:
$$\varphi_{1}^{(0)}(V)=\frac{I(S=0)\left\{\mu_1^{(0)}(X)-\psi^{(0)}_{1}\right\}+\sum_{k=0}^K\frac{C_{2k}}{\sum_{j=0}^{K}C_{2j}}H_2^{(k)}(V)}{\operatorname{pr}(S=0)}.$$

\subsection{Proof of Theorem 3}
In this section, we denote the EIFs in Theorems 1 and 2 by $\varphi^{(0)}_{1,\text{con1}}$ and $\varphi^{(0)}_{1,\text{con2}}$, respectively. When Condition $2^{(1:K)}$ holds, i.e., $\mu^{(0)}_a(X)=\cdots=\mu_a^{(K)}(X):=\mu_a(X)$ for $a=0,1$, we define $\tau(X)=\mu_1(X)/\mu_0(X)$. In addition, let $R_{1k}=C_{1k}/(\sum_{j=1}^{K}C_{1j})$ and $R_{2k}=C_{2k}/(\sum_{j=1}^{K}C_{2j})$ for $k\in[[K]]$. Then for the EIF in Theorem 2, we have

\begin{align*}
    \mathbb E[\operatorname{pr}(S=0)\varphi^{(0)}_{1,\text{con2}}\}^2] &=\mathbb E\left[I(S=0)\{\mu_1(X)-\psi_1^{(0)}\}^2\right]\\
    &+\mathbb E\left[\frac{2R_{20}I(S=0)\{\mu_1(X)-\psi_1^{(0)}\}A\{Y-\mu_{1}(X)\}}{\pi^{(0)}(X)}\right]\\
    &+\sum_{k=0}^K\mathbb E\left[\frac{R_{2k}^2I(S=k)A[q^{(k)}(X)\{Y-\mu_1(X)\}]^2}{\{\pi^{(k)}(X)\}^2}\right]\\
    &= \mathbb E\left[I(S=0)\{\mu_1(X)-\psi_1^{(0)}\}^2\right]\\
    &+\mathbb E\left[2R_{2k}\frac{\mu_1(X)-\psi_1^{(0)}}{\pi^{(0)}(X)}\mathbb E\left[\left.I(S=0)A\{Y-\mu_{1}(X)\}\right|X\right]\right]\\
    &+\sum_{k=0}^K\mathbb E\left[\frac{R_{2k}^2\{q^{(k)}(X)\}^2}{\{\pi^{(k)}(X)\}^2}\mathbb E[I(S=k)A\{Y-\mu_1(X)\}^2|X]\right]\\
    &=\mathbb E\left[I(S=0)\{\mu_1(X)-\psi_1^{(0)}\}^2\right]+\sum_{k=0}^K\mathbb E\left[\frac{R_{2k}^2q^{(k)}(X)\operatorname{pr}(S=0|X)\sigma_{1,k}^2(X)}{\pi^{(k)}(X)}\right].
\end{align*}

For the EIF in Theorem 1, we have

\begin{align*}
    H_1^{(k)}(V)&=\frac{I(S=k)Aq^{(k)}(X)\{Y-\mu_{1}(X)\}}{\pi^{(k)}(X)}\\
    &-\frac{I(S=k)(1-A)q^{(k)}(X)\tau(X)\{Y-\mu_{0}(X)\}}{1-\pi^{(k)}(X)}\\
    &+\frac{I(S=0)(1-A)\tau(X)\{Y-\mu_{0}(X)\}}{1-\pi^{(0)}(X)}.
\end{align*}
Then,
\begin{align*}
    \mathbb E[\{\operatorname{pr}(S=0)\varphi^{(0)}_{1,\text{con1}}\}^2] &= \mathbb E[\mathbb E[\{\operatorname{pr}(S=0)\varphi^{(0)}_{1,\text{con1}}\}^2|X]]\\
    &=\mathbb E\left[I(S=0)\{\mu_1(X)-\psi_1^{(0)}\}^2\right]\\
    &+\mathbb E\left[\frac{2R_{10}I(S=0)\{\mu_1(X)-\psi_1^{(0)}\}Aq^{(k)}(X)\{Y-\mu_{1}(X)\}}{\pi^{(k)}(X)}\right]\\
    &+\mathbb E\left[\sum_{k=1}^K\frac{2R_{1k}I(S=0)\{\mu_1(X)-\psi_1^{(0)}\}(1-A)\tau(X)\{Y-\mu_{0}(X)\}}{1-\pi^{(0)}(X)}\right]\\
    &+\sum_{k=0}^K\mathbb E\left[\frac{R_{1k}^2I(S=k)A[q^{(k)}(X)\{Y-\mu_1(X)\}]^2}{\{\pi^{(k)}(X)\}^2}\right]\\
    &+\sum_{k=0}^K\mathbb E\left[\frac{R_{1k}^2I(S=k)(1-A)[q^{(k)}(X)\tau(X)\{Y-\mu_0(X)\}]^2}{\{1-\pi^{(k)}(X)\}^2}\right]\\
    &+\sum_{k=0}^K\mathbb E\left[R_{1k}^2\frac{I(S=0)(1-A)[\tau(X)\{Y-\mu_0(X)\}]^2}{\{1-\pi^{(0)}(X)\}^2}\right]\\
    &+\sum_{i\neq j \& i,j\neq0}\mathbb E\left[R_{1i}R_{1j}\frac{I(S=0)(1-A)[\tau(X)\{Y-\mu_0(X)\}^2]}{\{1-\pi^{(0)}(X)\}^2}\right]\\
    &=\mathbb E\left[I(S=0)\{\mu_1(X)-\psi_1^{(0)}\}^2\right]\\
    &+\sum_{k=0}^K\mathbb E\left[\frac{R_{1k}^2\operatorname{pr}(S=0|X)q^{(k)}(X)\sigma_{1,k}^2(X)}{\pi^{(k)}(X)}\right]\\
    &+\sum_{k=0}^K\mathbb E\left[\frac{R_{1k}^2\operatorname{pr}(S=0|X)q^{(k)}(X)\{\tau(X)\}^2\sigma_{0,k}^2(X)}{1-\pi^{(k)}(X)}\right]\\
    &+\sum_{k=0}^K\mathbb E\left[R_{1k}^2\frac{\operatorname{pr}(S=0|X)\{\tau(X)\}^2\sigma_{0,0}^2(X)}{1-\pi^{(0)}(X)}\right]\\
    &+\sum_{i\neq j \& i,j\neq0}\mathbb E\left[R_{1i}R_{1j}\frac{\operatorname{pr}(S=0|X)\{\tau(X)\}^2\sigma_{0,0}^2(X)}{1-\pi^{(0)}(X)}\right]\\
    &=\mathbb E\left[I(S=0)\{\mu_1(X)-\psi_1^{(0)}\}^2\right]\\
    &+\sum_{k=0}^K\mathbb E\left[\frac{R_{1k}^2\operatorname{pr}(S=0|X)q^{(k)}(X)\sigma_{1,k}^2(X)}{\pi^{(k)}(X)}\right]\\
    &+\sum_{k=0}^K\mathbb E\left[\frac{R_{1k}^2\operatorname{pr}(S=0|X)q^{(k)}(X)\{\tau(X)\}^2\sigma_{0,k}^2(X)}{1-\pi^{(k)}(X)}\right]\\
    &+\mathbb E\left[\frac{\{(1-R_{10})^2+R_{10}^2\}\operatorname{pr}(S=0|X)\{\tau(X)\}^2\sigma_{0,0}^2(X)}{1-\pi^{(0)}(X)}\right]\\
    &\geq \mathbb E\left[I(S=0)\{\mu_1(X)-\psi_1^{(0)}\}^2\right]+\sum_{k=0}^K\mathbb E\left[\frac{R_{1k}^2q^{(k)}(X)\operatorname{pr}(S=0|X)\sigma_{1,k}^2(X)}{\pi^{(k)}(X)}\right].
\end{align*}

Since the estimator lies in the tangent space and thus attains the semiparametric efficiency bound under Condition $2^{(1:K)}$ only when the weights are given by $R_{2k}$, $k\in[[K]]$, it follows that $\mathbb E[\{\varphi^{(0)}_{1,\text{con1}}\}^2]\geq\mathbb E[\{\varphi^{(0)}_{1,\text{con2}}\}^2]$.

For the EIF of $\psi^{(0)}$, denoted by $\varphi_{\text{con1}}^{(0)}$ under Condition $1^{(1:K)}$ and $\varphi_{\text{con2}}^{(0)}$ under Condition $2^{(1:K)}$ respectively, we adopt a geometric approach that characterizes the EIFs through their corresponding tangent spaces.

For Condition $2^{(1:K)}$, the restrictive part of the tangent space for $\psi^{(0)}$ is:
\begin{align*}
    \mathcal T_{42,\text{con2}}&=\cap_{k\neq 0}	\left\{d(Y,X,S,A):\mathbb{E}\left\{\left[\frac{I(S=0)AY}{\operatorname{pr}(S=0|X)\pi^{(0)}(X)}-\frac{I(S=k)AY}{\operatorname{pr}(S=k|X)\pi^{(k)}(X)}\right]d(Y,X,S,A)\bigg|X\right\}=0\right\}\\
    &\cap_{k\neq 0}	\left\{d(Y,X,S,A):\mathbb{E}\left\{\left[\frac{I(S=0)(1-A)Y}{\operatorname{pr}(S=0|X)\{1-\pi^{(0)}(X)\}}-\frac{I(S=k)(1-A)Y}{\operatorname{pr}(S=k|X)\{1-\pi^{(k)}(X)\}}\right]d(Y,X,S,A)\bigg|X\right\}=0\right\}.
\end{align*}

For Condition $1^{(1:K)}$, when Condition $2^{(1:K)}$ holds, the restrictive part of the tangent space for $\psi^{(0)}$ is $\mathcal T_{42,\text{con1}}=\left\{d(Y,X,S,A): \mathbb{E}\{\eta_1^{(k)}(V)d(Y,X,S,A)|X\}=0\right\}$,
where
\begin{align*}
    \eta_1^{(k)}(V)&=\frac{I(S=0)A}{\operatorname{pr}(S=0|X)\pi^{(0)}(X)}\frac{Y}{\mu_{0}(X)}-\frac{I(S=0)(1-A)}{\operatorname{pr}(S=0|X)\{1-\pi^{(0)}(X)\}}\frac{Y\tau(X)}{\mu_{0}(X)}\\
&-\frac{I(S=k)A}{\operatorname{pr}(S=k|X)\pi^{(k)}(X)}\frac{Y}{\mu_{0}(X)}+\frac{I(S=k)(1-A)}{\operatorname{pr}(S=k|X)\{1-\pi^{(k)}(X)\}}\frac{Y\tau(X)}{\mu_{0}(X)}.
\end{align*}

For any $d^*(Y,X,S,A)\in\mathcal T_{42,\text{con2}}$, it satisfies
\begin{align*}
    &\cap_{k\neq 0}	\left\{d(Y,X,S,A):\mathbb{E}\left\{\left[\frac{I(S=0)AY}{\operatorname{pr}(S=0|X)\pi^{(0)}(X)}-\frac{I(S=k)AY}{\operatorname{pr}(S=k|X)\pi^{(k)}(X)}\right]\frac{d^*(Y,X,S,A)}{\mu_0(X)}\bigg|X\right\}=0\right\}\\
    &\cap_{k\neq 0}	\left\{d(Y,X,S,A):\mathbb{E}\left\{\left[\frac{I(S=0)(1-A)Y}{\operatorname{pr}(S=0|X)\{1-\pi^{(0)}(X)\}}-\frac{I(S=k)(1-A)Y}{\operatorname{pr}(S=k|X)\{1-\pi^{(k)}(X)\}}\right]\frac{d^*(Y,X,S,A)\tau(X)}{\mu_0(X)}\bigg|X\right\}=0\right\}.
\end{align*}

Therefore, 
\begin{align*}
    d^*(Y,X,S,A)\in\left\{d(Y,X,S,A): \mathbb{E}\{\eta_1^{(k)}(V)d(Y,X,S,A)|X\}=0\right\}=\mathcal T_{42,\text{con1}}.
\end{align*}

Hence, $\mathcal T_{42,\text{con2}}\subset\mathcal T_{42,\text{con1}}$. We know that the semiparametric efficiency bound equals the variance of the EIF, which is obtained as the orthogonal projection of any regular influence function onto the tangent space of the model. Therefore, under Condition $2^{(1:K)}$, only the projection onto $\mathcal T_{42,\text{con2}}$ achieves the corresponding semiparametric bound. Consequently, the semiparametric efficiency bound for $\psi^{(0)}$ corresponding to the EIF in Theorem 1 is larger than that for the EIF in Theorem 2 when Condition $2^{(1:K)}$ holds.

\subsection{Proof of Theorem 4 and Theorem 6 (\romannumeral1)}
\begin{assumption}
\label{regu}
    Let $V_i=(Y_i,X_i,A_i,S_i)$, the following regularity conditions hold:
    \begin{enumerate}
        \item[\romannumeral1)] $\varphi_1^{(0)}$ belongs to a Donsker class \citep{van2000asymptotic}.
        \item[\romannumeral2)] $\varphi_1^{(0)}$ is differentiable with respect to $\psi_1^{(0)}$ and $\mathbb E\{\partial\varphi_1^{(0)}(V)/\partial\psi_1^{(0)}\}$ exists and is invertible.
    \end{enumerate}
\end{assumption}

Following the empirical process literature, we denote $\widehat{\mathbb{P}}$ as the empirical measure over the data set across all sites, i.e., $\widehat{\mathbb{P}}\{h(V)\}=n^{-1}\sum_{i=1}^nh(V_i)$. Let $\mathbb{P}$ denotes the expectation over the data generative distribution, i.e., $\mathbb{P}\{h(V)\}=\int h(V)d\mathbb{P}$. Under Assumption \ref{regu}, by standard Taylor expansion, we have
\begin{align*}
    \widehat\psi_1^{(0)}-\psi_1^{(0)} &=-\left[\mathbb E\left\{\frac{\partial\varphi_1^{(0)}(V;\psi_{1}^{(0)},\{\pi^{(k)},\mu_0^{(k)}\}_{k=0}^K,\{q^{(k)}\}_{k=1}^K,\tau)}{\psi_1^{(0)}}\right\}\right]^{-1}\\
    &\times\widehat{\mathbb P}\{\varphi_1^{(0)}(V;\psi_{1}^{(0)},\{\widehat\pi^{(k)},\widehat\mu_0^{(k)}\}_{k=0}^K,\{\widehat q^{(k)}\}_{k=1}^K,\widehat\tau)\}+o_p(n^{-1/2})\\
    &=\widehat{\mathbb P}\{\varphi_1^{(0)}(V;\psi_{1}^{(0)},\{\widehat\pi^{(k)},\widehat\mu_0^{(k)}\}_{k=0}^K,\{\widehat q^{(k)}\}_{k=1}^K,\widehat\tau)\}+o_p(n^{-1/2})
\end{align*}

Moreover, we can show
\begin{align}
    &\widehat{\mathbb{P}}\left\{\varphi_{1}^{(0)}\left(V;\psi_{1}^{(0)},\{\widehat\pi^{(k)},\widehat\mu_0^{(k)}\}_{k=0}^K,\{\widehat q^{(k)}\}_{k=1}^K,\widehat\tau\right)\right\}\notag\\ 
    &=\widehat{\mathbb{P}}\left\{\varphi_{1}^{(0)}\left(V;\psi_{1}^{(0)},\{\pi^{(k)},\mu_0^{(k)}\}_{k=0}^K,\{q^{(k)}\}_{k=1}^K,\tau\right)\right\}\notag\\
    \label{2term}&+\mathbb{P}\left\{\varphi_{1}^{(0)}\left(V;\psi_{1}^{(0)},\{\widehat\pi^{(k)},\widehat\mu_0^{(k)}\}_{k=0}^K,\{\widehat q^{(k)}\}_{k=1}^K,\widehat\tau\right)-\varphi_{1}^{(0)}\left(V;\psi_{1}^{(0)},\{\pi^{(k)},\mu_0^{(k)}\}_{k=0}^K,\{ q^{(k)}\}_{k=1}^K,\tau\right)\right\}\\
    \label{3term}&+\left(\widehat{\mathbb{P}}-\mathbb{P}\right)\left\{\varphi_{1}^{(0)}\left(V;\psi_{1}^{(0)},\{\widehat\pi^{(k)},\widehat\mu_0^{(k)}\}_{k=0}^K,\{\widehat q^{(k)}\}_{k=1}^K,\widehat\tau\right)-\varphi_{1}^{(0)}\left(V;\psi_{1}^{(0)},\{\pi^{(k)},\mu_0^{(k)}\}_{k=0}^K,\{q^{(k)}\}_{k=1}^K,\tau\right)\right\}
\end{align}
where the third term (\ref{3term}) is $o_p(n^{-1/2})$ under \ref{regu} \romannumeral1). Even if the Donsker condition in Assumption \ref{regu} \romannumeral2) is not met, the cross-fitting procedure in \citet{chernozhukov2018double} can be used to assure that (\ref{3term}) is negligible. We now show that the second term (\ref{2term}) is a small order term:

\begin{align*}
    &\operatorname{pr}(S=0)\mathbb{P}\left\{\varphi_{1}^{(0)}\left(V;\psi_{1}^{(0)},\{\widehat\pi^{(k)},\widehat\mu_0^{(k)}\}_{k=0}^K,\{\widehat q^{(k)}\}_{k=1}^K,\widehat\tau\right)-\varphi_{1}^{(0)}\left(V;\psi_{1}^{(0)},\{\pi^{(k)},\mu_0^{(k)}\}_{k=0}^K,\{ q^{(k)}\}_{k=1}^K,\tau\right)\right\}\\
    &=\sum_{k=0}^K\mathbb{P}\left\{\widehat{R}_{1k}\frac{\operatorname{pr}(S=k|X)\pi^{(k)}(X)\widehat{q}^{(k)}(X)}{\widehat{\pi}^{(k)}(X)}\frac{\{\mu_{0}^{(k)}(X)\tau(X)-\widehat{\mu}_{0}^{(k)}(X)\widehat{\tau}(X)\}\widehat{\mu}_{0}^{(0)}(X)}{\widehat{\mu}_{0}^{(k)}(X)}\right\}\\
    &-\sum_{k=0}^K\mathbb{P}\left\{\widehat{R}_{1k}\frac{\operatorname{pr}(S=k|X)(1-\pi^{(k)}(X))\widehat{q}^{(k)}(X)}{1-\widehat{\pi}^{(k)}(X)}\frac{\{\mu_{0}^{(k)}(X)-\widehat{\mu}_{0}^{(k)}(X)\}\widehat{\tau}(X)\widehat{\mu}_{0}^{(0)}(X)}{\widehat{\mu}_{0}^{(k)}(X)}\right\}\\
    &+\sum_{k=0}^K\mathbb{P}\left\{\widehat{R}_{1k}\frac{\operatorname{pr}(S=k|X)(1-\pi^{(0)}(X))q^{(k)}(X)}{1-\widehat{\pi}^{(0)}(X)}\{\mu_{0}^{(0)}(X)-\widehat{\mu}_{0}^{(0)}(X)\}\widehat{\tau}(X)\right\}\\
    &-\sum_{k=0}^K\mathbb{P}\left\{\widehat{R}_{1k}\operatorname{pr}(S=k|X)q^{(k)}(X)(\mu_{0}^{(0)}(X)\tau(X)-\widehat{\mu}_{0}^{(0)}(X)\widehat{\tau}(X))\right\}\\
    &=\sum_{k=0}^K\mathbb{P}\left\{\widehat{R}_{1k}\operatorname{pr}(S=k|X)\frac{\widehat{q}^{(k)}(X)\widehat{\tau}(X)\widehat{\mu}_{0}^{(0)}(X)}{\widehat{\mu}_{0}^{(k)}(X)}\frac{\{\widehat{\pi}^{(k)}(X)-\pi^{(k)}(X)\}\{\widehat{\mu}_{0}^{(k)}(X)-\mu_{0}^{(k)}(X)\}}{\widehat{\pi}^{(k)}(X)(1-\widehat{\pi}^{(k)}(X))}\right\}\\
    &+\sum_{k=0}^K\mathbb{P}\left\{\widehat{R}_{1k}\operatorname{pr}(S=k|X)\frac{\widehat{q}^{(k)}(X)\widehat{\mu}_{0}^{(0)}(X)\pi^{(k)}(X)\{\mu_{0}^{(k)}(X)-\widehat{\mu}_{0}^{(k)}(X)\}\{\tau(X)-\widehat{\tau}(X)\}}{\widehat{\mu}_{0}^{(k)}(X)\widehat{\pi}^{(k)}(X)}\right\}\\
    &+\sum_{k=0}^K\mathbb{P}\left\{\widehat{R}_{1k}\operatorname{pr}(S=k|X)\frac{\widehat{q}^{(k)}(X)\widehat{\mu}_{0}^{(0)}(X)\{\pi^{(k)}(X)-\widehat{\pi}^{(k)}(X)\}\{\tau(X)-\widehat{\tau}(X)\}}{\widehat{\pi}^{(k)}(X)}\right\}\\
    &+\sum_{k=0}^K\mathbb{P}\left\{\widehat{R}_{1k}q^{(k)}(X)\operatorname{pr}(S=k|X)\widehat{\tau}(X)\frac{\{\pi^{(0)}(X)-\widehat{\pi}^{(0)}(X)\}\{\widehat{\mu}_{0}^{(0)}(X)-\mu_{0}^{(0)}(X)\}}{1-\widehat{\pi}^{(0)}(X)}\right\}
\end{align*}

From the result above, under Assumption \ref{con1regul}, the $\pi^{(k)}(X)$, $\mu_0^{(k)}(X)$, $\tau(X)$ and $q^{(k)}(X)$ for $k\in[[K]]$ is uniformly bounded, then by Cauchy Inequality, the above term is bounded by 
\begin{align*}
&\sum_{k=0}^Ka_{1k}\|\widehat{\pi}^{(k)}(X)-\pi^{(k)}(X)\|\times\|\widehat{\mu}_{0}^{(k)}(X)-\mu_{0}^{(k)}(X)\|\\
&+\sum_{k=0}^Ka_{2k}\|\widehat{\pi}^{(k)}(X)-\pi^{(k)}(X)\|\times\|\widehat{\tau}(X)-\tau(X)\|\\
&+\sum_{k=0}^Ka_{3k}\|\widehat{\mu}_{0}^{(k)}(X)-\mu_{0}^{(k)}(X)\|\times\|\widehat{\tau}(X)-\tau(X)\|\\
&+\sum_{k=1}a_{4k}\|\widehat{q}^{(k)}(X)-q^{(k)}(X)\|\times\|\widehat{\tau}(X)-\tau(X)\|,
\end{align*}
which is $o_p(n^{-1/2})$. Therefore, the bias is asymptotically negligible, and the EIF can be used for obtaining a robust variance estimation.

\subsection{Proof of Theorem 5 and Theorem 6 (\romannumeral2)}
We now show that the second term (\ref{2term}) is a small order term when Condition $2^{(1:K)}$ holds:

\begin{align*}
&\operatorname{pr}(S=0)\mathbb{P}\left\{\varphi_{1}^{(0)}\left(V;\psi_{1}^{(0)},\{\widehat\pi^{(k)}\}_{k=0}^K,\{\widehat q^{(k)}\}_{k=1}^K,\widehat\mu_1\right)-\varphi_{1}^{(0)}\left(V;\psi_{1}^{(0)},\{\pi^{(k)}\}_{k=0}^K,\{q^{(k)}\}_{k=1}^K,\mu_1\right)\right\}\\
&=\mathbb{P}\left\{\sum_{k=0}^{K}\widehat{R}_{2k}q^{(k)}(X)\operatorname{pr}(S=k|X)\{\widehat{\mu}_{1}(X)-\mu_{1}(X)\}\right\}\\
&+\mathbb{P}\left\{\sum_{k=0}^{K}\widehat{R}_{2k}\frac{\operatorname{pr}(S=k|X)\pi^{(k)}(X)\widehat{q}^{(k)}(X)\{\mu_{1}(X)-\widehat{\mu}_{1}(X)\}}{\widehat{\pi}^{(k)}(X)}\right\}\\
&=\mathbb{P}\left\{\sum_{k=0}^{K}\widehat{R}_{2k}\operatorname{pr}(S=k|X)\{\widehat\mu_{1}(X)-{\mu}_{1}(X)\}\left(q^{(k)}(X)-\frac{\pi^{(k)}(X)\widehat{q}^{(k)}(X)}{\widehat{\pi}^{(k)}(X)}\right)\right\}\\
&=\mathbb{P}\left\{\sum_{k=0}^{K}\widehat{R}_{2k}\operatorname{pr}(S=k|X)\{\widehat\mu_{1}(X)-{\mu}_{1}(X)\}\left(q^{(k)}(X)-\widehat{q}^{(k)}(X)+\frac{\{\widehat{\pi}^{(k)}(X)-\pi^{(k)}(X)\}\widehat{q}^{(k)}(X)}{\widehat{\pi}^{(k)}(X)}\right)\right\}\\
&=\mathbb{P}\left\{\sum_{k=0}^{K}\widehat{R}_{2k}\operatorname{pr}(S=k|X)\{\widehat\mu_{1}(X)-{\mu}_{1}(X)\}\{q^{(k)}(X)-\widehat{q}^{(k)}(X)\}\right\}\\
&+\mathbb{P}\left\{\sum_{k=0}^{K}\widehat{R}_{2k}\operatorname{pr}(S=k|X)\{\widehat\mu_{1}(X)-\mu_{1}(X)\}\frac{\{\widehat{\pi}^{(k)}(X)-\pi^{(k)}(X)\}\widehat{q}^{(k)}(X)}{\widehat{\pi}^{(k)}(X)}\right\}
\end{align*}

From the result above, under Assumption \ref{con2regul}, the $\pi^{(k)}(X)$, $q^{(k)}(X)$ and $\mu_1(X)$ for $k\in[[K]]$ are uniformly bounded, then by Cauchy Inequality, the above term is bounded by 
\begin{align*}
&\sum_{k=0}^Kb_{1k}\|\widehat{\pi}^{(k)}(X)-\pi^{(k)}(X)\|\times\|\widehat{\mu}_{1}(X)-\mu_{1}(X)\|\\
&+\sum_{k=1}^Kb_{2k}\|\widehat{q}^{(k)}(X)-q^{(k)}(X)\|\times\|\widehat{\mu}_{1}(X)-\mu_{1}(X)\|,
\end{align*}
which is $o_p(n^{-1/2})$ under Assumption \ref{con2regul}. Therefore, the bias is asymptotically negligible, and the EIF can be used for obtaining a robust variance estimation.

\subsection{Proof of Theorem 7}
The argument follows the oracle property proof structure for adaptive penalized estimators, inspired by the proof of Lemma 4 in \citet{han2025federated}. The key logic proceeds in two steps: (\romannumeral1) First, we establish that $\|\widetilde{\pmb w}-\overline{\pmb w}\|=O_p(n^{-1/2})$ by showing that the objective functions of the optimization problems defining $\widetilde{\pmb w}$ and $\overline{\pmb w}$ differ only by $O_p(n^{-1/2})$; (\romannumeral2) Second, we show that $\widetilde{\pmb w}=\widehat{\pmb w}$ with high probability, which ensures the oracle selection set $\mathcal S$ can be correctly chosen asymptotically.

We define the oracle selection space for $\pmb w$ as
$$\mathcal S=\{k\in[[K]]:\tau^{(k)}(x)=\tau^{(0)}(x)\},\quad \mathbb R^{\mathcal S}=\{\pmb w\in\mathbb R^{K+1}:w_k=0,k\notin\mathcal S\},$$
and the asymptotic loss function as
$$L^*(\pmb w)=\mathbb{P}_{n}\left\{\left[\varphi^{(0)<0>}_1\left(V;{\psi}^{(0)<0>}_1,{\pi}^{(0)},{\mu}_1^{(0)}\right)-\sum_{k\neq0}w_{k}\varphi^{(0)<k>}_1\left(V;{\psi}^{(0)<0>}_1, \{\pi^{(l)},\mu_0^{(l)}\}_{l=0,k},q^{(k)},\tau\right)\right]^{2}\right\}.$$
Correspondingly, define its empirical version as

$$\widehat L(\pmb w)=\mathbb{P}_{n}\left\{\left[\varphi^{(0)<0>}_1\left(V;\widehat{\psi}^{(0)<0>}_1,\widehat{\pi}^{(0)},\widehat{\mu}_1^{(0)}\right)-\sum_{k\neq 0}w_{k}\varphi^{(0)<k>}_1\left(V;\widehat{\psi}^{(0)<0>}_1, \{\widehat\pi^{(l)},\widehat\mu_0^{(l)}\}_{l=0,k},\widehat q^{(k)},\widehat\tau\right)\right]^{2}\right\}.$$

Then we denote the follwing minimizers:
$$\overline{\pmb w}={\arg\min}_{\pmb w\in\mathbb R^{\mathcal S}}L^*(\pmb w),$$
$$\widetilde{\pmb w}={\arg\min}_{\pmb w\in\mathbb R^{\mathcal S}}\widehat L(\pmb w)+\lambda_n\sum_{k\neq 0}|w_k|\left(\widehat\delta^{(k)}\right)^2,$$
and
$$\widehat{\pmb w}={\arg\min}_{\pmb w\in\mathbb R^{K+1}}\widehat L(\pmb w)+\lambda_n\sum_{k\neq 0}|w_k|\left(\widehat\delta^{(k)}\right)^2.$$

Under Assumption 4, we have $|\widehat L(\pmb w)-L^*(\pmb w)|=O( n^{-1/2})$. We know for $k\in\mathcal S$, $\widehat\delta^{(k)}=|\widehat\psi^{(0)<k>}_1-\widehat\psi^{(0)<0>}_1|=O_p(n^{-1/2}).$
Given $n^{-1/2}\lambda_n\rightarrow0$, the penalty term is asymptotically negligible in a compact neighborhood of $\overline{\pmb w}$, i.e.,
$$\sup_{\|\pmb w-\overline{\pmb w}\|\leq M}\lambda_n\sum_{k\neq 0}|w_k|\left(\widehat\delta^{(k)}\right)^2=O_p(n^{-1/2}).$$

Hence,
$$\sup_{\|\pmb w-\overline{\pmb w}\|\leq M}\left|\widehat L(\pmb w)+\lambda_n\sum_{k\neq 0}|w_k|\left(\widehat\delta^{(k)}\right)^2-L^*(\pmb w)\right|=O_p(n^{-1/2}).$$

By the convexity of the optimization problem, it follows that $\|\widetilde{\pmb w}-\overline{\pmb w}\|=O_p(n^{-1/2})$.

The optimality condition of the original problem is
$$
\left\{
\begin{array}{cc}
    \frac{\partial}{\partial w_k}\widehat L=-\text{sign}(w_k)\lambda_n(\widehat\delta^{(k)})^2, & w_k\neq0, \\
    \left|\frac{\partial}{\partial w_k}\widehat L\right|\leq\lambda_n(\widehat\delta^{(k)})^2, & w_k=0.
\end{array}
\right.
$$

For $k\in\mathcal S$, i.e., $w_k\neq0$, $\widetilde{\pmb w}$ satisfies these conditions by definition. To establish the optimality of $\widetilde{\pmb w}$, it suffices to verify that
\begin{equation}
\label{optcon}
    \left|\frac{\partial}{\partial w_k}\widehat L\right|\leq\lambda_n\left(\widehat\delta^{(k)}\right)^2,\quad k\in[[K]]\backslash\mathcal S.
\end{equation}
By the definition of $\mathcal S$, we have for biased sites $\delta^{(k)}\neq0$. Therefore, we have for $k\in[[K]]\backslash\mathcal S$, $\widehat\delta^{(k)}=\delta^{(k)}+O_p(n^{-1/2})$, which is bounded away from zero. With $n^{(\gamma-1)/2}\lambda_n\rightarrow\infty$, the penalty for biased sites diverges for $k\in[[K]]\backslash\mathcal S$, i.e., $\lambda_n(\widehat\delta^{(k)})^2\rightarrow\infty$.

Thus, condition \ref{optcon} holds with high probability, implying that $\widetilde{\pmb w}$ satisfies the exact optimality condition with high probability. By the convexity of the problem, we conclude that $\widehat{\pmb w}=\widetilde{\pmb w}$ with high probability. Thus we have, $\|\overline{\pmb w}-\widehat{\pmb w}\|=O_p(n^{-1/2})$, which means that $\widehat w_k\xrightarrow{p} 0$, for $k\notin \mathcal S$.

\section{Data generating process of simulation studies}
\label{dgp}
The data for the simulation study were generated as follows. We conduct $M=500$ Monte Carlo replications with a total of $n=\sum_{k=0}^Kn_k$ observations distributed across $3$ sites. We fix the total sample size at $n=1000$. Each
site $k\in\{0, 1, 2\}$ contains $n_k$ observations. The site indicator variable $S$ is generated with probabilities $f(S=0)=0.1$, $f(S=1)=0.4$ and $f(S=2)= 0.5$. The covariate $X\in\mathbb R$ follows a normal distribution within each site, i.e., $X|S=0\sim \mathcal N(2, 1)$, $X|S=1\sim \mathcal N(1,4)$ and $X|S=2\sim \mathcal N(2,4)$.

Next, we define the propensity score function $\pi^{(k)}$, the conditional outcome function $\mu_0^{(k)}$ and the conditional RR function $\tau^{(k)}$ for $k\in[[2]]$ as follows:
$$\pi^{(0)}(X)=\frac{1}{1+\exp(1-0.5x)},\quad\mu_{0}^{(0)}(x)=x,\quad\tau^{(0)}(x)=x;$$
$$\pi^{(1)}(x)=\frac{1}{1+\exp(1-0.8x)},\quad\mu_{0}^{(1)}(x)=x+b^{(1)}_\mu(x),\quad\tau^{(1)}(x)=x+b^{(1)}_\tau(x);$$
$$\pi^{(2)}(x)=\frac{1}{1+\exp(1-0.3x)},\quad\mu_{0}^{(2)}(x)=x+b^{(2)}_\mu(x),\quad\tau^{(2)}(x)=x+b^{(2)}_\tau(x),$$
where $b_\mu^{(k)}(x)$ and $b_\tau^{(k)}$ vary across different scenarios described in the main text.

Finally, the potential outcomes in each site are generated as $(Y(1)|X,S=k)\sim \mathcal N\left(\mu_0^{(k)}(X)\tau^{(k)}(X),1\right)$$
$$(Y(0)|X,S=k)\sim \mathcal N\left(\mu_0^{(k)}(X),1\right)$ for $k\in[[2]]$. The observed outcome is then given by $Y=AY(1)+(1-A)Y(0)$.

The data generating process above supports the scenarios introduced in $\S$5.1.

\section{Additional real data diagnostic}
\label{realdata}
\begin{figure}[htbp]
    \centering
    \includegraphics[width=0.8\linewidth]{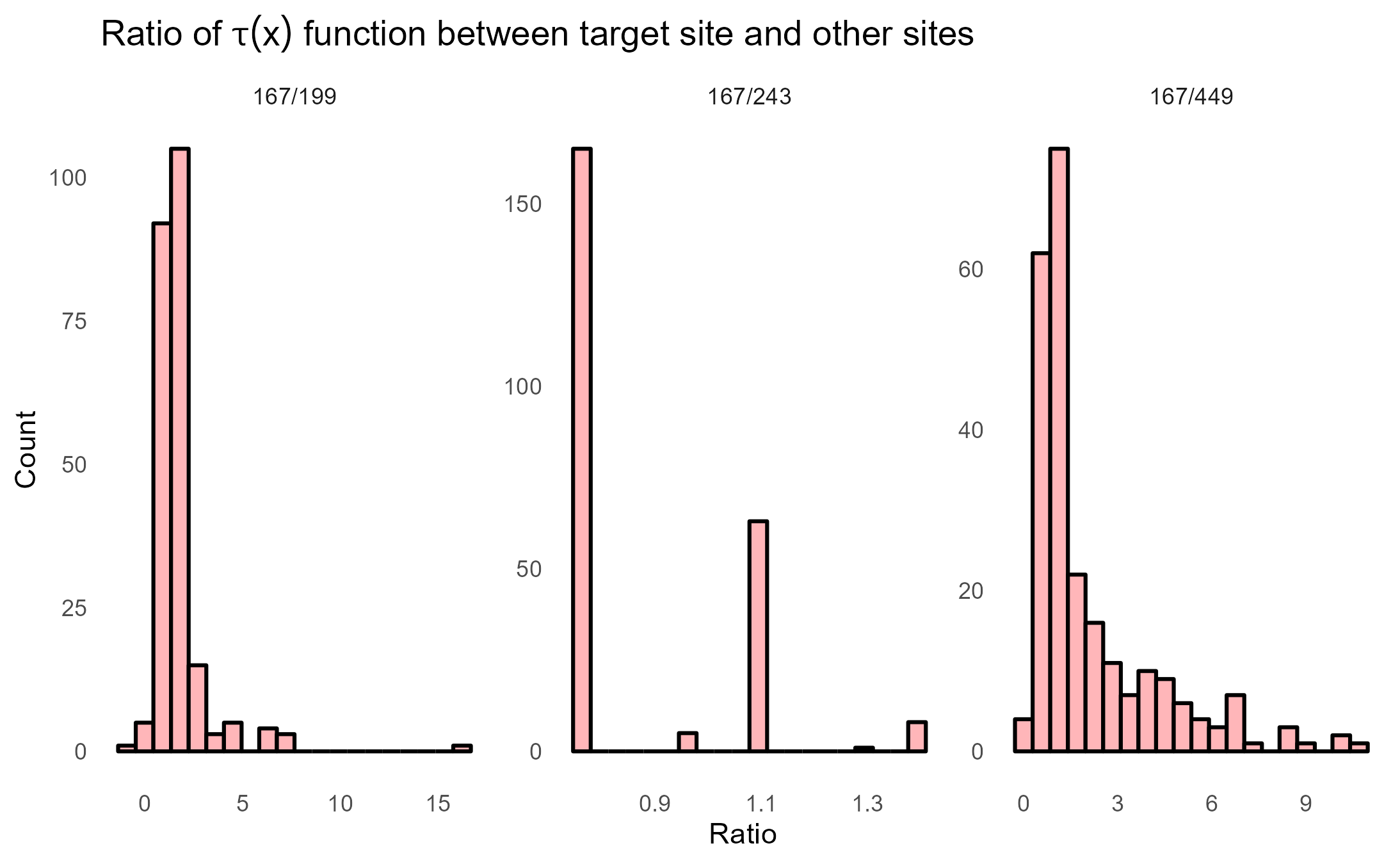}
    \caption{Ratio of $\widehat{\tau}(X)$ between the target site and each source site, computed using the pooled data from the two sites.}
    \label{taur}
\end{figure}

To provide a supplementary diagnostic of cross-site heterogeneity, we compare the estimated treatment effect functions $\tau(x)$ between the target site and each source site. Specifically, for each covariate vector $x$, we compute the ratio $\widehat\tau^{(167)}(x)/\widehat\tau^{(k)}(x)$ for $k=199, 243, 449$. As shown in Figure \ref{taur}, only site 243 exhibits a $\tau(x)$ function similar to that of the target site, indicating that our method correctly selects the site most likely to satisfy Condition $1^{(k)}$.

\end{document}